**Syngas conversion to higher alcohols via wood-framed Cu/Co-carbon catalyst**

*Guihua Yan[1,2,3], Paulina Pršlja[4], Gaofeng Chen[1,3], Jiahui Kang[2], Yongde Liu[1], Miguel A. Caro[4], Xi Chen[2,5], Xianhai Zeng[3]*, Bo Peng[2,6]**

[1] College of Environmental Engineering, Henan University of Technology, 450001, China

[2] Department of Applied Physics, Aalto University, FI-00076 Aalto, Finland

[3] College of Energy, Xiamen University, 361102, China

[4] Department of Chemistry and Materials Science, Aalto University, Kemistintie 1, 02150 Espoo, Finland

[5] Present address: The School of Physical Science and Technology, Lanzhou University, 730000, China

[6] Department of Materials Science, Advanced Coatings Research Center of Ministry of Education of China, Fudan University, 200433, China

E-mails: xianhai.zeng@xmu.edu.cn, peng_bo@fudan.edu.cn

Guihua Yan, Paulina Pršlja, and Gaofeng Chen contributed equally.

## Abstract

Syngas conversion into higher alcohols represents a promising avenue for transforming coal or biomass into liquid fuels. However, the commercialization of this process has been hindered by the high cost, low activity, and inadequate $C_{2+}OH$ selectivity of catalysts. Herein, we have developed Cu/Co carbon wood catalysts, offering a cost-effective and stable alternative with exceptional selectivity for catalytic conversion. The formation of Cu/Co nanoparticles was found, influenced by water-1,2-propylene glycol ratios in the solution, resulting in bidisperse nanoparticles. The catalyst exhibited a remarkable CO conversion rate of 74.8% and a selectivity of 58.7% for $C_{2+}OH$, primarily comprising linear primary alcohols. This catalyst demonstrated enduring stability and selectivity under industrial conditions, maintaining its efficacy for up to 350 h of operation. We also employed density functional theory (DFT) to analyze selectivity, particularly focusing on the binding strength of CO, a crucial precursor for subsequent reactions leading to the formation of $CH_3OH$. DFT identified the pathway of $CH_x$ and CO coupling, ultimately yielding $C_2H_5OH$. This computational understanding, coupled with high performance of the Cu/Co-carbon wood catalyst, paves ways for the development of catalytically selective materials tailored for higher alcohols production, thereby ushering in new possibility in this field.

**Keywords**: Syngas conversion, Cu/Co carbon wood catalyst, methanol, ethanol, density functional theory

# Introduction

Carbon neutrality stands as a paramount objective in contemporary efforts towards energy conservation and emission reduction.[1] The global environmental crisis has intensified the urgency to develop renewable energy sources.[2, 3] Among the alternatives to fossil fuels, higher alcohols synthesized from syngas, with more than two carbons emerge, are emerging as promising candidates because of their high octane numbers.[4, 5] Importantly, these higher alcohols can be derived from various sources including biomass, coal, natural, and shale gas, establishing a recognized pathway within the environmentally sustainable carbon cycle.[6-8]

Efficient catalysts are crucial for producing higher alcohols from syngas by striking a balance between C-C coupling and C-O retention. A variety of catalysts have been employed to facilitate this process, including Rh/Mo-based, methanol-synthesis, and Fischer-Tropsch-synthesized catalysts. Transition metals such as Cu, Co, Ni, and Fe are among the most prevalent utilized for this purpose.[9, 10] However, despite their widerspread use, challenges such as high support costs and catalyst agglomeration and deactivation induced by granular supports with intricate channels hinder industrial applications.[4, 6] In contrast, wood features clear and continues channels from top to bottom, exhibiting highly aligned and porous structure at both micro and nanoscales. This offers possibilities for structural and compositional design and modification.[11-13]

Over the past decade, advances in the physicochemical modification of wood has broadened its functional scope across diverse applications, particularly in fluidic and ionic transport.[14, 15] The abundance of chemical groups and anisotropic conduits in wood make it conducive decorating with various nanoparticles.[15, 16] Carbonized wood (CW) derived from natural wood, which is abundant, sustainable, and inexpensive inherits its straight channels and hierarchically porous structure from its origin form, making it ideal conduits for gas or ion transport.[17-20]

Here, we develop a wood-framed Cu/Co catalyst that demonstrates exceptional selectivity in converting syngas into higher alcohols. As depicted in **Figure 1**, the process begins by immersing natural wood in a solution of catalyst precursors, facilitating the capture of precursors thanks to the wood's abundant chemical groups. Subsequent thermal treatment leads to the formation of Cu/Co-based nanoparticles, and the carbonization of the wood, yielding Cu/Co-based nanoparticles embedded within a carbonized wood matrix, termed Cu/Co-CW. These Cu/Co-CW composites serve as catalysts for converting syngas into alcohols with remarkable selectivity. To elucidate the catalytic selectivity, we employ Density Functional Theory (DFT) analysis, employing CO binding energy as a descriptor, as CO is the key intermediate in higher alcohol production. We further investigate the mechanism following CO-adsorption to differentiate between *CHOH or $*OCH_2$ intermediates, eventually leading to $CH_3OH$ production. Additionally, we explore the coupling of $CH_3$ and CO to produce $CH_3CH_2OH$. This simple yet effective synthesis strategy, coupled with computational insights into selectivity, offers a promising avenue for fabricating a 3D structured catalyst for syngas conversion into higher alcohols.

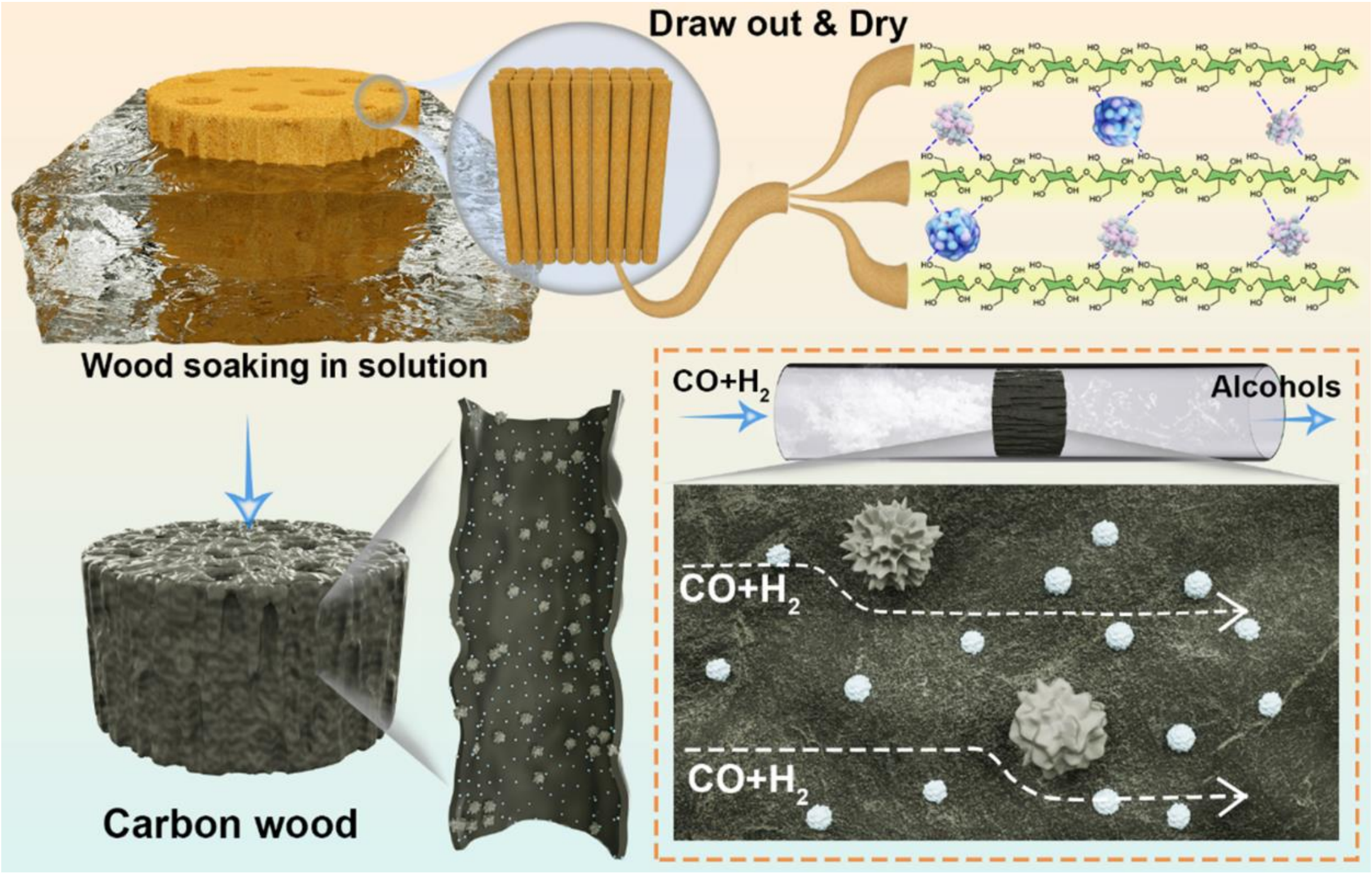

**Figure 1**. **Fabrication scheme of an internally crosslinked carbonized wood hierarchical structure.** After carbonization, the aligned 3D structure is retained with two sizes of nanoparticles uniformly dispersed within the wood channels. As syngas flows through the carbon channels, catalytic reaction occurs, leading to alcohol production.

## Results and discussion

### Synthesis and structural characterization of the Cu/Co-CW catalysts

The abundance of chemical groups in wood facilitates the easy decoration of various nanoparticles. The anisotropic conduit serves as a confined and directed reactor, enabling streamlined production and mitigating nanoparticles aggregation, thereby elevating the catalytic efficiency. The synthesis of the Cu/Co carbon wood catalysts is detailed in the experimental section under the methodology.

Upon carbonization, the brown natural wood transitions to black carbonized wood (see **Figure 2a**). The resulting Cu/Co-CW maintains open channels as its primary skeleton structure (**Figure 2b**), preserving the original porous architecture of natural wood (**Figure S1**). **Figures 2b-j** present scanning electron microscopy (SEM) results of the Cu/Co-CW catalyst at different viewing angles and magnifications. Cross-sectional SEM images reveal channels with diameters ranging from 5 to 60 μm, hosting uniformly distributed nanoparticles approximately 200 nm in diameter on their inner walls (**Figures 2c-e**). Closer observation reveals smaller nanoparticles deposited within the carbon walls (**Figure 2f**). Additionally, these nanoparticles are uniformly dispersed along the wood channels (see **Figures 2g-h**). Further magnification shows a subset of smaller nanoparticles (~ 15-30 nm) uniformly distributed around the larger particles and adhered to the inner surface of the carbonized wood, indicating the formation of bidisperse nanoparticles (**Figures 2i and 2j**). These bidisperse nanoparticles are present not only on the surface but also within the wood channels, underscoring the efficiency of our synthesis strategy. The uniform distribution of bidisperse nanoparticles is further corroborated by Energy Dispersive X-Ray Analysis (EDX), where elemental mapping images of C, O, Cu, and Co reveal distinct nanoparticle composites (**Figures 2k-l**).

In contrast, the Cu/Co-CW impregnated with deionized water as a solvent demonstrates uniform nanoparticle loading (~20 nm) on the wood walls (see **Figure S2**). These nanoparticles contain both Cu and Co elements, indicating the formation of nano-alloy (**Figures S3-S5**). Furthermore, the Brunauer-Emmett-Teller data reveal that the Cu/Co-CW possesses a substantial surface area and pore volume (**Figure S6**). Previous reports suggest that this distinctive tubular channel structure facilitates the rapid passage of gases and liquids.[21]

In addition to deionized water, 1,2-propanediol is used as the co-solvent to promote the formation of nanoparticles. The ratio between water and 1,2-propanediol varies from 1:0, to 0.5:0.5 and further to 0.33:0.67, and the resulting catalysts are named Cu/Co-CW-W, Cu/Co-CW-$W_1P_1$, and Cu/Co-CW-$W_1P_2$, respectively. The Cu/Co-CW-$W_1P_1$ catalyst exhibits the highest Cu/Co content among three samples and the control samples (see **Figure S7**). This likely because 1,2-propanediol can form complex with precursors, resulting in the formation of different nanoparticle types after sintering.

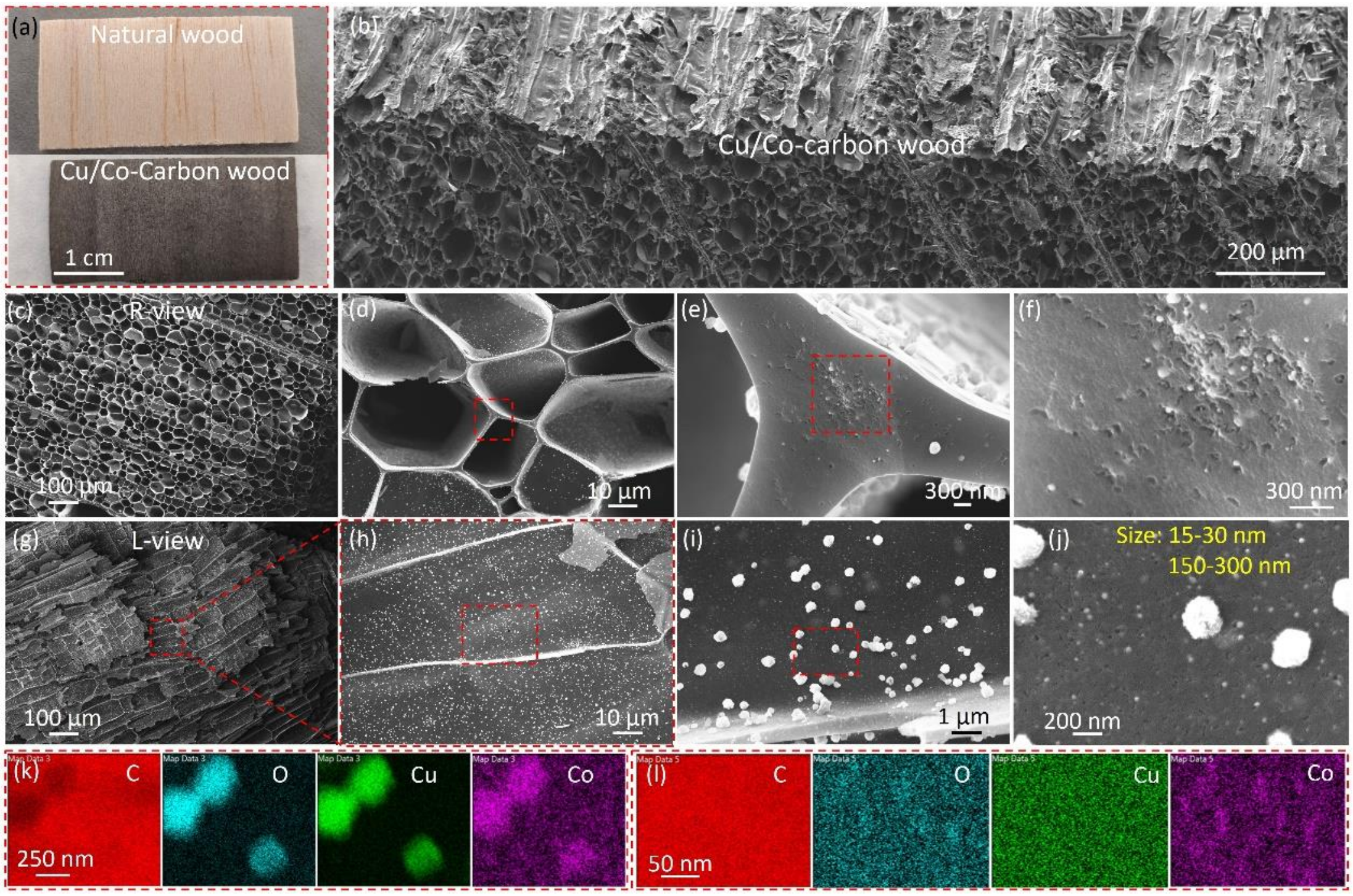

**Figure 2**. **Morphological characterization of Co/Cu-CW-$W_1P_1$**. (a) The photos of natural wood and corresponding carbonized wood. SEM images of (b) Cu/Co-CW-$W_1P_1$, (c-d) Cu/Co-CW-$W_1P_1$ in the radial direction (R-direction) and (e-f)s corresponding enlarged view. SEM images of (g) Cu/Co-CW-$W_1P_1$ in the longitudinal direction (L-direction), and (h-j) corresponding enlarged views. Elemental mapping images of (k, l) C, O, Co, and Cu distributed in Cu/Co-CW-$W_1P_1$ at different magnifications.

Furthermore, we characterize the distribution of nanoparticles with high-resolution transmission electron microscopy (HR-TEM). The TEM images clearly show the uniform growth of bi-disperse particles on the surface of CW after calcination (see **Figure 3a**). The uniform distribution hints that the 1,2-propanediol/deionized water mixture can effectively permeate into the wood channels, facilitating the homogenously absorption of $Cu^{2+}$/$Co^{2+}$ precursors by woods and then uniform nanoparticle formation (see **Figures 3b, 3c, and S8**). In addition, the sizes of bidisperse nanoparticles are 15-30 nm and 150-300 nm (**Figure S9)**, consistent with the SEM observation. Chemical composition analysis of the marked area in **Figure 3b** confirms the presence of C, O, Cu, and Co through energy-dispersive X-ray spectroscopy (EDX) elemental analysis. Notably, small nanoparticles predominantly contain the Co element, while the large nanoparticles primarily consist of Cu, O, and a minor Co content (**Figures S10** and **S11**). HR-TEM and TEM-EDX analyses (**Figure 3d**) reveal that the small nanoparticles compose $Co_2C$, exhibiting interplanar spacing of approximately 2.08 nm, 2.18 nm, 2.42 nm, and 2.43 nm, corresponding to the (021), (020), (101) and (101) planes of cobalt carbide ($Co_2C$). Conversely, due to their large size, no distinct interplanar spacings are observable in clusters mainly composed of Cu and O (**Figure S11**).

For modeling the Cu/Co-CW-$W_1P_1$ catalyst, we utilize $Co_2C$ (101, 020) surfaces to represent smaller clusters, while CuO(111), Cu(100) and Cu(111) surfaces to represent larger particles (**Figure 3e**). Although Cu-based planes are not experimentally identified, surfaces known to produce methanol[22-26] and ethanol[9, 27, 28] are modeled accordingly. SEM and corresponding EDX analysis of Cu/Co-CW-W and Cu/Co-CW-$W_1P_2$ (**Figures S12** and **S13**) reveal monodisperse Cu/Co nanoparticles and bidisperse nanoparticles in the case of Cu/Co-CW-$W_1P_2$ catalyst. This suggests that the addition of 1,2-propanediol as a solution influence the creation of nanoparticles with varying sizes. 1,2-propanediol can form complex with precursors, resulting in the formation of different types of nanoparticles after sintering.

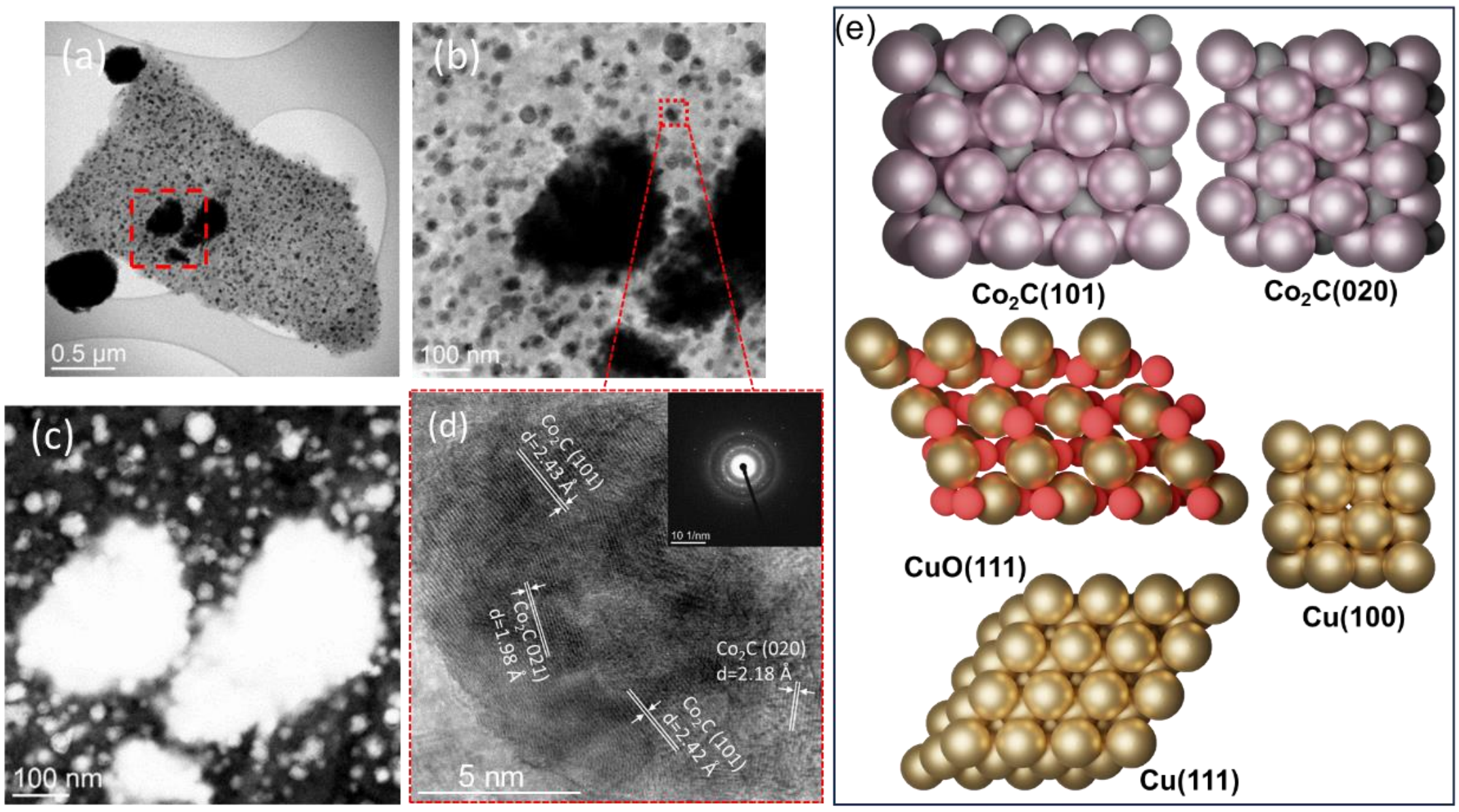

**Figure 3**. **Morphological characterization of the Cu/Co-CW-$W_1P_1$ catalyst.** (a) Cs-

corrected HR-TEM, its corresponding (b) enlarged view, and (c) HAADF-STEM image. (d) High-resolution TEM image. (e) Optimized computational models of the $Co_2C$, CuO, and Cu surfaces (top view).

The chemical structure of the Cu/Co-CW-$W_1P_1$ nanoparticles are analyzed with X-ray diffraction (XRD) spectra (**Figure 4a**). The peaks at 28.6°, 37.1°, 40.2°, and 45.7° can be attributed to the (011), (110), (020) and (002) planes of $Co_2C$ (PDF#72-1369). Additionally, peaks at 44.2° and 51.1° are mainly attributed to Co metal (PDF#89-4307), while peaks at 35.4° and 38.9° correspond to CuO (PDF#89-5895), and the peaks at 43.3°, 50.4°, and 73.9° correspond to metallic Cu (PDF#85-1326). These XRD results well align with the TEM analysis. Moreover, XRD patterns for CW, Cu/Co-CW-W, and Cu/Co-CW-$W_1P_2$ are analyzed (see **Figure S14**). The absence of characteristic $Co_2C$ peaks in CW and Cu/Co-CW-W confirms that the 1,2-propanediol/deionized water mixture promotes the formation of $Co_2C$ clusters, unlike the aqueous solution. The broad XRD signal of CW indicates the presence of amorphous carbon material.

Raman spectra reveal significant changes in the carbon samples upon the incorporation of Cu/Co nanoparticles (see **Figure S15**). Specifically, the Raman spectrum of Cu/Co-CW-based samples exhibits two characteristic peaks: The D band (~1334 $cm^{-1}$), attributed to sp3 hybridized carbon, and the G band (~1586 $cm^{-1}$), corresponding to the sp2 hybridized carbon atoms and carbon lattice defects.[29] The ratio of $I_D/I_G$ for the CW, Cu/Co-CW-$W_1P_1$, Cu/Co-CW-$W_1P_2$, and Cu/Co-CW-W are measured as 0.91, 0.97, 1.00, and 0.99, respectively, indicating a high density of structural defects introduced by the presence of Cu/Co.

Next, the chemical states of the C, O, Cu, and Co elements are confirmed by X-ray photoelectron spectroscopy (XPS). The full XPS pattern of Cu/Co-CW-$W_1P_1$ reveals the presence of C, O, Cu, and Co elements (see **Figure S16**). In the C1s XPS spectrum, peaks are centered at 287.1, 285.3, 284.6, and 283.7 eV, corresponding to C=O, C-O, C-C, and Co-C bonding, respectively (see **Figure 4b**). Furthermore, the Co 2p spectrum shows the presence of Co primarily as $Co_2C$ and Co metal, with $2p_{3/2}$ peaks at around 785.8, 781.9, and 778.6 eV, respectively (see **Figure 4c**). The presence of $Co_2C$ is confirmed by the C 1s and Co 2p spectra, validating the composition of small particles as $Co_2C$ and Co metal.

The photoelectron and auger electron peaks of Cu element show the existence of Cu primarily as $Cu^{2+}$ and $Cu^0$, with $2p_{3/2}$ peaks centered at 934.5 eV and 933.3 eV, corresponding to CuO and Cu metal, respectively (see **Figure 4d** and **Figure S17**). In addition, the XPS spectra of CW, Cu/Co-CW-W, and Cu/Co-CW-$W_1P_2$ catalysts are analyzed (**Figures S18** and **S19**). Cu/Co-CW-$W_1P_2$ exhibits characteristic $Co_2C$ peaks, while Cu/Co-CW-W does not, indicating that the addition of 1,2-propanediol to the mixture can form complexes with precursors, resulting in different types of nanoparticles after sintering, which play an important role in the formation of bidisperse nanoparticles.

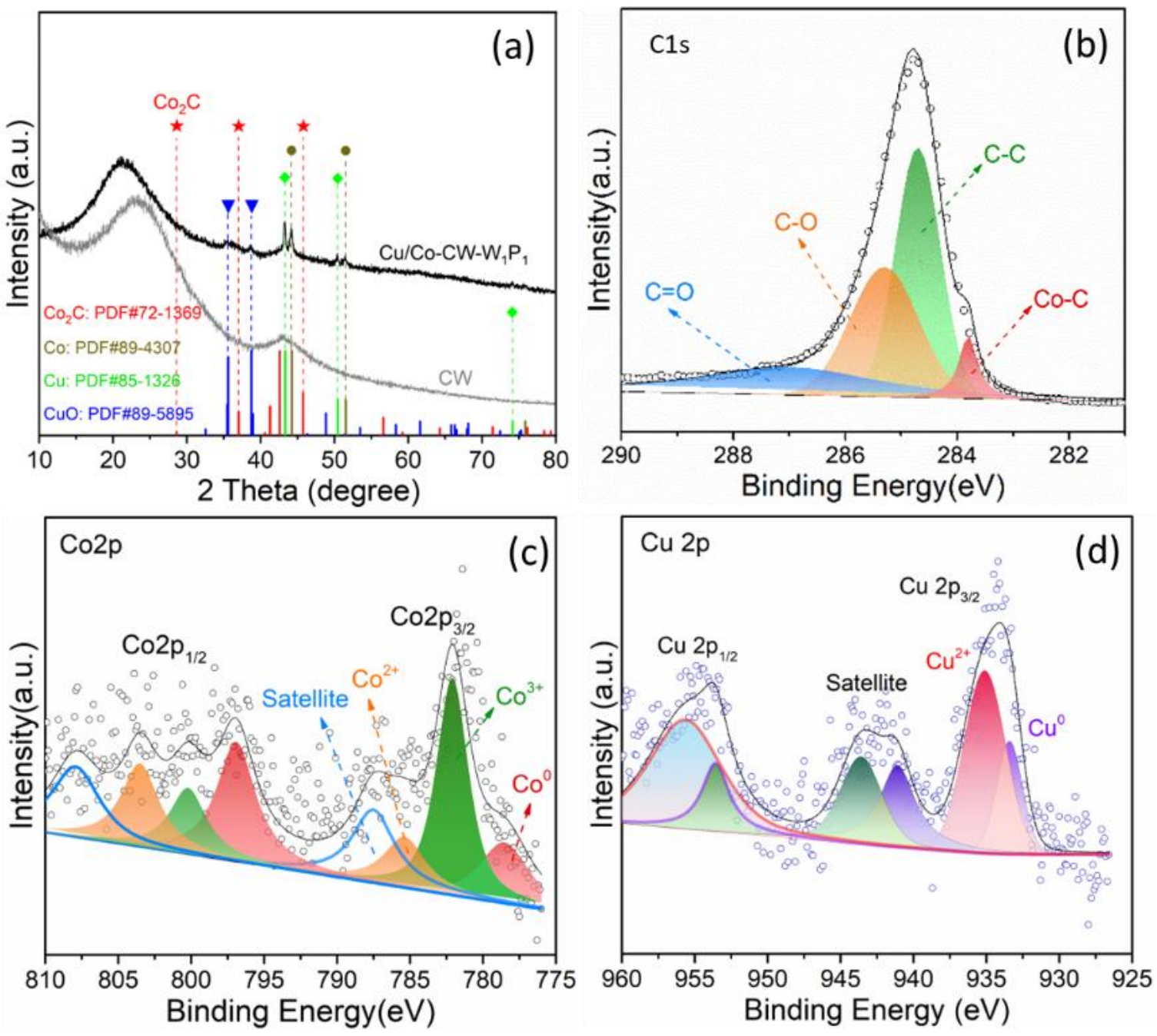

**Figure 4. Chemical composition analysis** of the Cu/Co-CW-$W_1P_1$ by (a) XRD spectra, and (b-d) XPS spectra of the C $1_s$, Co $2_p$, and Cu $2_p$.

## Mechanism exploration of syngas conversion

Herein, we use DFT insights to elucidate the formation mechanism of $CH_3OH$ and $C_2H_5OH$ from syngas. The initial phase of the reaction involves the simultaneous adsorption of CO and H, leading to the reorganization into *CHO intermediates. Alternatively, if CO is adsorbed independently, it undergoes hydrogenation to form *CHO.

Subsequently, *CHO can undergo further hydrogenation, yielding two possible intermediates: formaldehyde *$CH_2O$ (see **Figure 5a**) or *CHOH (**Figure 5b**). The hydrogenation of *$OCH_2$ leads to the formation of *$OCH_3$, while *CHOH evolves into *$CH_2OH$, both contributing to the production of $CH_3OH$. Alternatively, initial *CO can be hydrogenated into *COH, which subsequently yields $CH_3OH$ through a *CHOH intermediate.

Furthermore, *$OCH_3$ can dissociate into *$CH_3$ and *O, where *$CH_3$ can be further hydrogenated to produce $CH_4$ (see **Figure 5c**). Dissociated *$CH_3$ and $CH_x$ species resulting from *C hydrogenation (see **Figure 5d**) can undergo coupling with CO, forming a C-C bond. This process involves CO insertion into $CH_3$ or $CH_2$, producing *$CH_3CO$ and *$CH_2CO$ intermediates.

The *$CH_3CO$ intermediate can undergo hydrogenation to form *$CH_3COH$ or *$CH_3CHO$, both of which proceed to form *$CH_3CHOH$, ultimately contributing to the production of $C_2H_5OH$ (see **Figure 5e)**. On the other hand, *$CH_2CO$ can be hydrogenated to *$CH_3CO$ or *$CH_2COH$. The hydrogenation of *$CH_3CO$ follows the pathway leading to ethanol production, shown in **Figure 5e**. Meanwhile, the hydrogenation of *$CH_2COH$ yields *$CH_3COH$ or *$CH_2CHOH$, both of which further react to form *$CH_3CHOH$. Notably *$CH_2CHOH$ can also transform into *$CH_2CH_2OH$, further contributing to the production of $C_2H_5OH$ (**Figure 5f**).

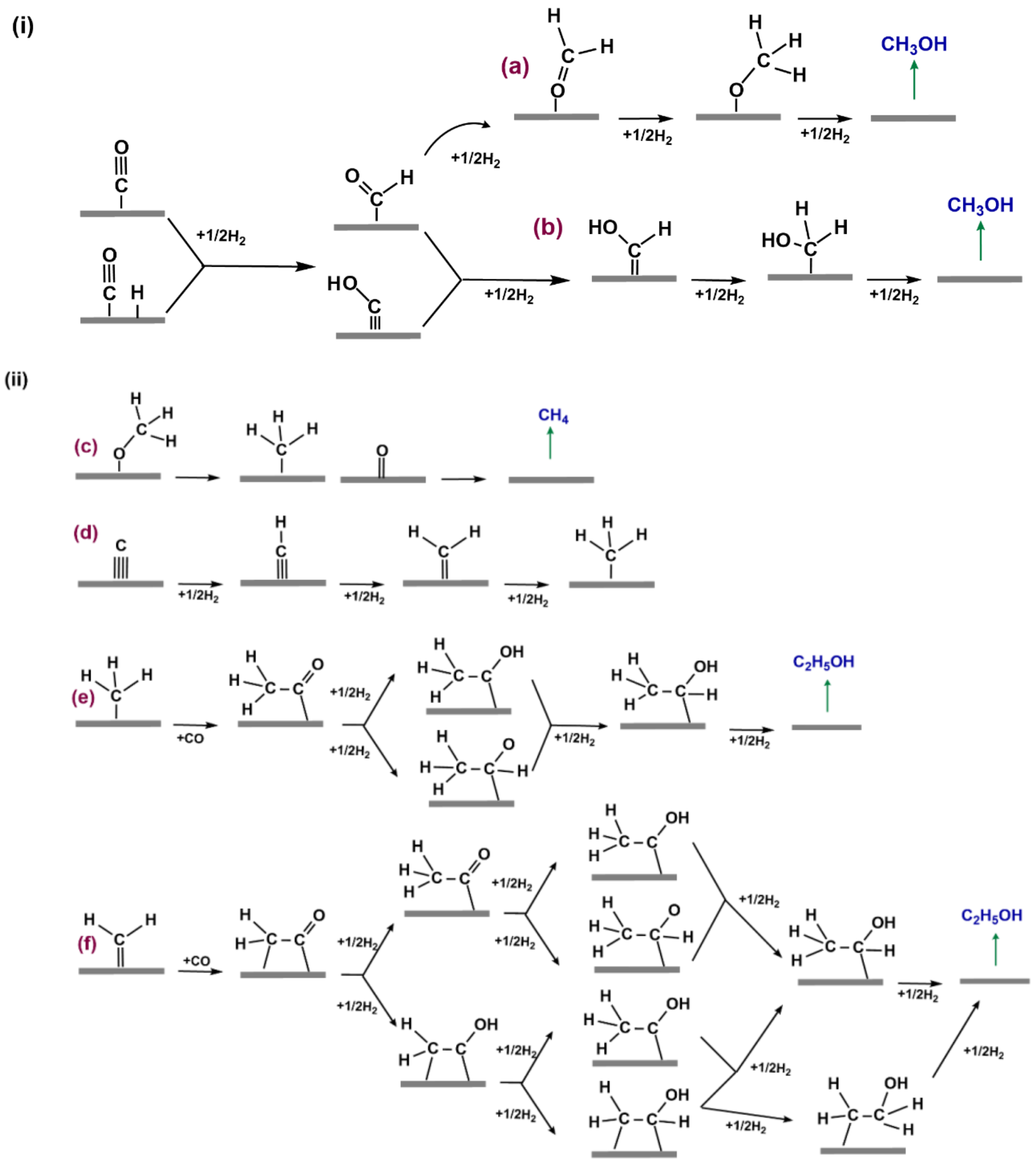

**Figure 5**. **Reaction network of syngas conversion.** Reaction mechanism toward (i) $CH_3OH$ and (ii) $C_2H_5OH$ production, where $CH_3OH$ can be produced via (a) $*OCH_2$ and (b) $*CHOH$ intermediates. The formation of $*CH_3$ occurs through (c) $*OCH_3$ dissociation and (d) $*C$ hydrogenation. $C_2H_5OH$ is formed via CO insertion into (e)$*CH_3$ or (f)$*CH_2$, where C-C bonding occurs.

### Catalytic performance of the Cu/Co-CW catalysts

The catalytic performance of the Cu/Co-CW composite is evaluated using the Micro-activity Effi reactor (see **Figure S20**) under the reaction conditions of 3 MPa, 270 °C, $H_2/CO = 2/1$, and gas hourly space velocity (GHSV) = 9600 mL $g_{cat}^{-1}$ $h^{-1}$. Strips of the 3D Cu/Co-CW- based catalysts are prepared by cutting along the growth direction of the tree and loaded into the reactor. The open 3D channels along the growth direction in the Cu/Co-CW facilitate the fast transport of the CO and $H_2$ mixture (see **Figure S21**). Compared to straight channels, the curved channels and variations in nanoparticle sizes within the Cu/Co-CW could slow down gas flow, promoting the contact between the mixture and the catalyst, thereby improving reaction efficiency.

**Figure 6** shows the product distribution over the studied catalysts. The Cu/Co-CW-$W_1P_1$ and Cu/Co-CW-$W_1P_2$ catalysts exhibit high CO conversion rates of 74.8 and 61.2%, respectively with alcohol selectivity of 52.3% and 42.9%. However, with the 1,2-propanediol/water ratio increases, both CO conversion and alcohol selectivity showed a declining trend. This suggests that the activation of CO and $H_2$ could be modulated by different sizes, influenced by the ratio of added 1,2-propanediol/water during the catalyst synthesis. Among the catalysts, the Cu/Co-CW-$W_1P_1$ catalyst displays the highest $C_{2+}OH$ selectivity of 30.7% and the lowest $CH_4$ and $CO_2$ selectivity of 26.1 and 2.7%, respectively (see **Figure 6a and Table S2**). Additionally, the Cu/Co-CW- $W_1P_1$ catalyst exhibits the highest ROH yield at 39.1%. The higher alcohols produced by this catalyst are mainly linear primary alcohols, as shown in **Table S3**.

**Figure 6b** depicts the $C_{2+}OH$ selectivity and ROH space time yields for the studied catalysts, revealing a volcano-like trend, with the maximum observed at 58.7% and 570 mg $g_{cat}^{-1}$ $h^{-1}$ for Cu/Co-CW-$W_1P_1$ catalyst. The product distribution is presented in **Figure 6c** with quasi-Anderson-Schulz-Flory (ASF) plots, characterized by the chain growth probability factor ($\alpha$), independent of the number of carbon atoms in the product molecule. The ROH distribution across all Cu/Co-CW catalysts conforms to the classical ASF model. Notably, the carbon chain growth probability ($\alpha$) of alcohols is calculated by fitting the ASF distribution, revealing a volcano-like slope with the increase of the water/1,2-propanediol ratio. The highest carbon chain growth probability observed for the Cu/Co-CW-$W_1P_1$ catalyst favors the production of $C_{2+}OH$.

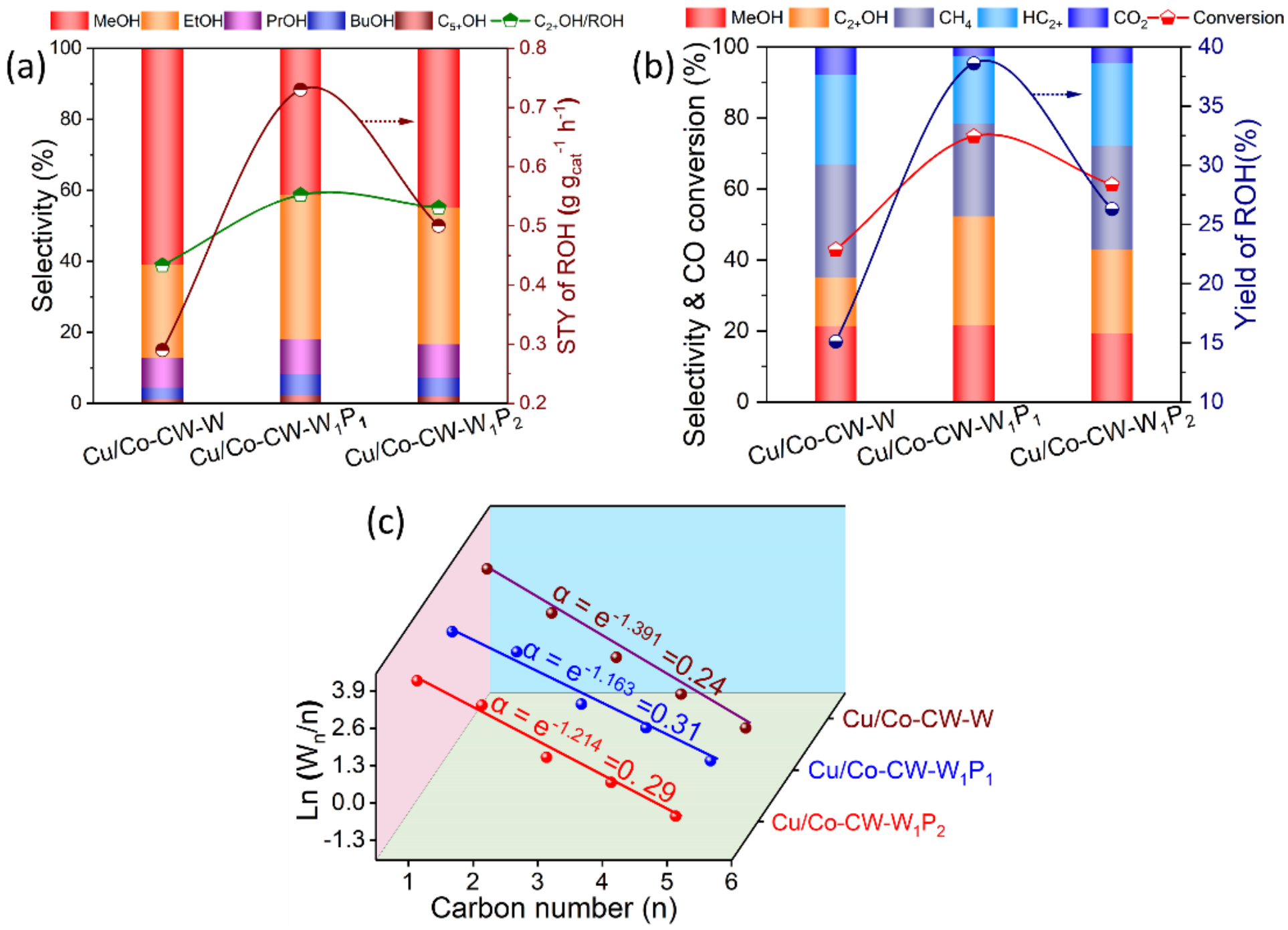

**Figure 6**. **The catalytic performance across all catalysts.** (a) CO hydrogenation conversion into ROH products and their yields. (b) The selectivity of alcohols and the space time yield of ROH. (c) Anderson-Schulz-Flory (ASF) plot.

We use the binding energy of CO as a selectivity descriptor, given its crucial roles as a key intermediate in the hydrogenation process leading to higher alcohols. According to our DFT calculations, the strength of CO binding energy follows a descending trend of -1.70 eV for $Co_2C$(020) > -1.57 eV for $Co_2C$(101) > -0.81 eV for CuO (111) > -0.42 eV for Cu(100) > -0.33 eV for Cu(111). This indicates that $Co_2C$ surfaces bind CO more strongly compared to Cu-based surfaces. To provide a computational reference, we introduce the CO binding energy over Rh(111), known for its production of $C_2H_5OH$,[30, 31] which is calculated to be -1.77 eV.[30] Therefore, the product's selectivity towards alcohol can be explained by the binding strength of CO.

Based on the obtained CO binding energy results, we conclude that the hydrogenation of CO is indeed feasible. Moreover, at a coverage of 0.3 monolayer (ML) on $Co_2C$(020) surface, CO preferably occupies a bridge site, while increasing the CO coverage favors a top site. Conversely, on the $Co_2C$(101), Cu(100), Cu(111), and CuO(111) surfaces at 0.3 ML coverage, CO predominantly occupies a top site. As the coverage increases to 0.5 ML and 0.8 ML, top sites are preferred for all studied surfaces.

Optimal Gibbs free energy diagrams over the studied models are shown in **Figure 7**. As previously mentioned, CO can undergo hydrogenation to form a formyl intermediate, *CHO, or a hydroxyl intermediate, *COH. However, on the $Co_2C$(101), Cu(100), and Cu(111) surfaces, the reaction proceeds through *CHO, as the energy barrier for *COH formation is higher by 0.79, 0.77, and 1.10 eV, respectively. On the $Co_2C$(020) surface, the energy barrier for *COH is lower by 0.03 eV compared to *CHO, while on CuO(111), *COH dissociates into *CO and *O. Additionally, *COH hydrogenation on $Co_2C$(020) leads to the disintegration of *CHOH into *CHO and *H, rendering the mechanism chemically implausible. CO hydrogenation continues via the *CHO intermediate on $Co_2C$(101), $Co_2C$(020)¸ Cu(100), Cu(111), and CuO(111) with energies of -3.87, 0.14, 0.57, 0.74, and 0.36 eV, respectively.

Furthermore, on the $Co_2C$(020), Cu(100), Cu(111), and CuO(111) surfaces, *CHO is hydrogenated into *$CH_2O$ with energies of 0.10, -0.14, -0.26, and 0.42 eV, respectively, while on $Co_2C$(101), it forms *CHOH with an energy of 0.30 eV. For $Co_2C$(101), the reaction proceeds through *$CH_2OH$ with an energy of -0.34 eV, while for $Co_2C$(020), Cu(100), Cu(111), and CuO(111), it processes via *$CH_3O$ with energies of -0.01, -0.81, -0.74, and 0.10 eV, respectively.

Considering the above, we infer that *CO, *CO*H, *CHO, *CHOH, *$CH_2OH$, *$CH_3OH$ (see **Figures 7a and S26, and Table S6**) is an optimal pathway for the initial CO hydrogenation on $Co_2C$(101), whereas on $Co_2C$(020), Cu(100), Cu(111) and CuO(111), the pathway follows *CO, *CO*H, *CHO, *$OCH_2$, *$OCH_3$, *$CH_3OH$ (see **Figures 7a 7c, and S26, and Table S7**). The divergence in reaction routes on $Co_2C$ surfaces can be attributed to the catalyst's oxophilicity, which correlates with the binding energy of O.[32, 33] $Co_2C$(101) binds *O stronger with -3.10 eV, while $Co_2C$(020) binds it with -2,68 eV. Consequently, *CHO hydrogenates to *CHOH on $Co_2C$(101) due to the stronger Co-O bond, making it harder to desorb.

Finally, the binding energy of $CH_3OH$ follows the trend of -4.38 eV for $Co_2C$(101) > 0.02 eV for CuO(111) > 0.15 eV for $Co_2C$(020) > 0.48 eV for Cu(100) > 0.50 eV for Cu(111). $CH_3OH$ gets poisoned on the $Co_2C$(101) surface as the C of the *$CH_2OH$ intermediate gets hydrogenated, leaving O to adsorb more strongly due to the oxophilicity of the $Co_2C(101)$ surface. Moreover, *O binds over $Co_2C(020)$ surface weaker, making the Co-O bond easier to break and desorb the $CH_3OH$. Conversely, Cu-based surfaces bind $CH_3OH$ significantly weaker, indicating that $CH_3OH$ would desorb due to the weak Cu-O bond.

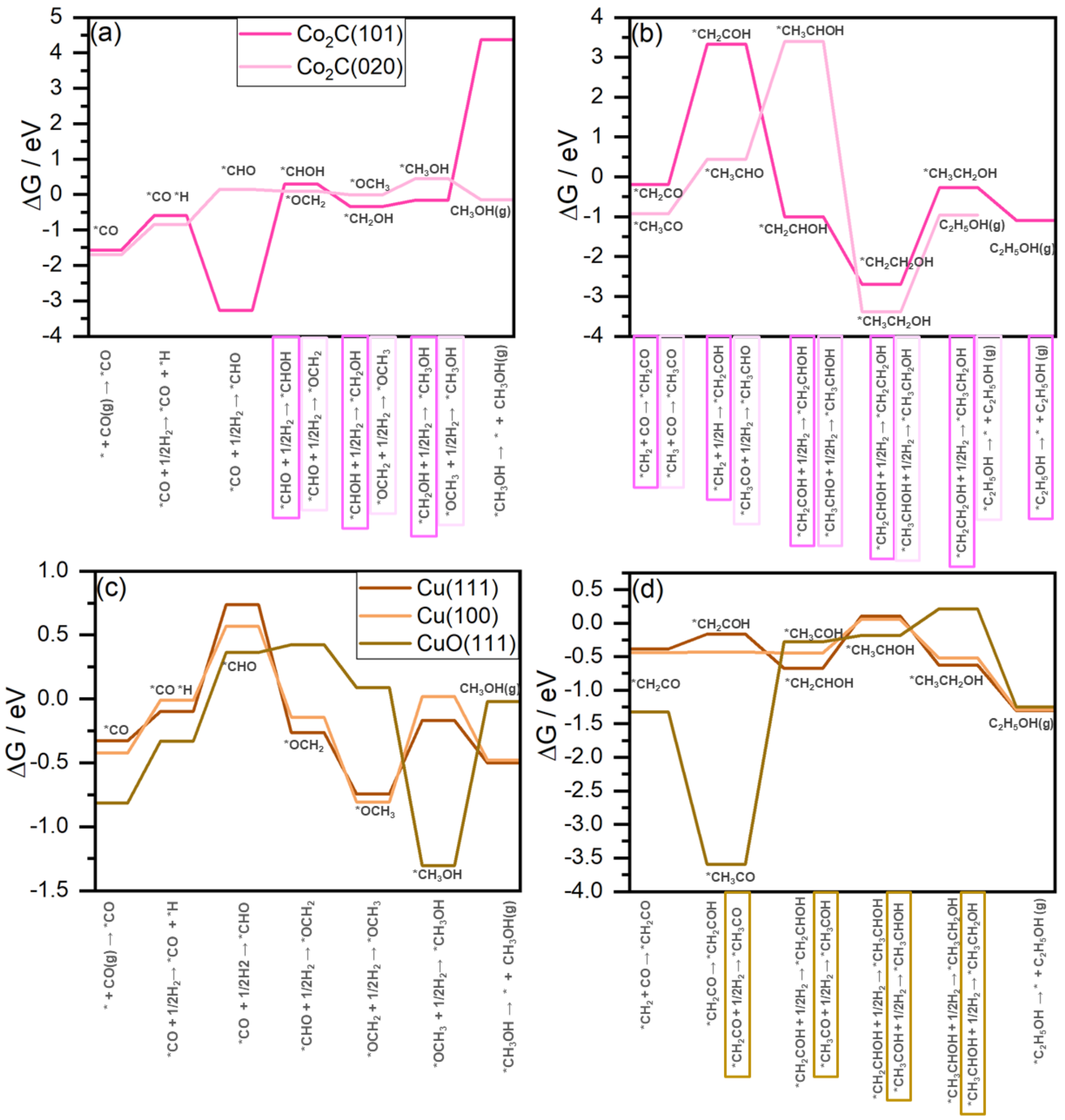

**Figure 7**. **Free Gibbs energy diagram** for (a, b) $Co_2C$ surfaces and (c, d) Cu-based surfaces toward (a, c) $CH_3OH$ and (b, d) $C_2H_5OH$ formation.

We explored ethanol formation through the coupling of $CH_x$ (x = 2 and 3) species and CO, a mechanism commonly discussed in connection with higher alcohol synthesis.[31, 34] $CH_x$ species can originates from *C hydrogenation (see **Figure 5d**), while $CH_3$ can also form from *$OCH_3$ dissociation (see **Figure 5c**). Dissociation of *$OCH_3$ is favorable on $Co_2C(020)$, and Cu(100) surfaces with energies of -0.78 and 0.07 eV, respectively. However, *$CH_3O$ cannot dissociate on $Co_2C(101)$, Cu(111), and CuO(111) surfaces due to energy barriers of 2.17, 0.43, and 0.74 eV, respectively.

Moreover, hydrogenation toward *$CH_2$ from CH can occur on all studied surfaces, with energy values of 0.16 eV for $Co_2C$(101), 0.10 eV for $Co_2C$(020), 0.22 eV for Cu(100), -0.56 eV for Cu(111), and -1.46 eV for CuO(111). It is noteworthy that the formation energies on $Co_2C$ surfaces and Cu(100) are slightly endothermic.

The direct insertion of CO into $CH_x$ species forms either *$CH_2CO$ or *$CH_3CO$. Based on $CH_x$-CO coupling, the reaction on $Co_2C$(101) starts with *$CH_2CO$ with a formation energy of -0.19 eV. Subsequent hydrogenation leads to *$CH_2COH$, *$CH_2CHOH$, *$CH_2CH_2OH$, *$CH_3CH_2OH$ with energies of 3.33, -1.01, -2,70, and -0.27 eV, respectively (see **Figures 7b and S27, and Table S8**). Conversely, on $Co_2C$(020), the reaction initiates from *$CH_3CO$ with a formation energy of -0.93 eV, followed by *$CH_3CHO$, *$CH_3CHOH$, *$CH_3CH_2OH$, with energies of 0.44, 3.39, and -3.38 eV, respectively (see **Figures 7b and S27, and Table S9**).

It is observed that the hydrogenation of $CH_2CO$ and $CH_3CO$ intermediates on $Co_2C$ surfaces is highly endothermic, indicating that the reaction may not proceed as predicted by DFT. Although ethanol production is experimentally observed on the Co/CW-$W_1P_1$ sample, we speculate that this reaction may be catalyzed by a different Co carbide surface, elemental Co, or edge/defect site. Carrying out an exhaustive exploration of all possible reactive sites at the DFT level is computationally impractical. Future work employing machine-learning-accelerated simulations may provide further insights into this issue.

On Cu(100), CO insertion into *$CH_2$ and *$CH_3$ generates *$CH_2CO$ and *$CH_3CO$ with energies of -0.44 and -0.39 eV, respectively. Notably, *$CH_2CO$ is thermodynamically more stable and proceeds to be hydrogenated into*$CH_2COH$, *$CH_2CHOH$, *$CH_3CHOH$, and *$CH_3CH_2OH$ with energies of -0.43, -0.44, 0.06, and -0.52 eV, respectively (see **Figures 7d and S27, and Table S8**). Similarly, on Cu(111), the reaction proceeds through hydrogenation of *$CH_2CO$ into *$CH_2COH$, *$CH_2CHOH$, *$CH_3CHOH$, and *$CH_3CH_2OH$ with energies of -0.39, -0.16, -0.68, 0.11, and -0.63 eV, respectively (see **Figures 7d and S27, Table S8**). On CuO(111), hydrogenation from *$CH_2CO$ continues into *$CH_3CO$, *$CH_3COH$, *$CH_3CHOH$, and *$CH_3CH_2OH$, with energies of -1.32, -3.59, -0.28, -0.18, and 0.21 eV, respectively (see **Figures 7d and S27, Table S9**). Finally, desorption energies for $CH_3CH_2OH$ follow the trend: -0.96 eV for $Co_2C$(020) < -1.10 eV for $Co_2C$(101) < -1.25 eV for CuO(111) < -1.29 eV for Cu(100) < -1.30 eV for Cu(111). Overall, the synthesis of $C_2H_5OH$ is feasible through direct CO insertion into $CH_x$ species.

## Stability of the Cu/Co-CW-$W_1P_1$ catalyst

To bolster industrial viability, ensuring the stability of the catalyst is paramount. Therefore, the lifespan of the Cu/Co-CW-$W_1P_1$ catalyst is investigated under operating conditions of 3 MPa pressure, 270 °C temperature, $H_2$/CO ratio of 2/1, and GHSV of 9600 mL $g_{cat}^{-1}$ $h^{-1}$ for 350 h. As shown in **Figure 8a**, CO conversion experiences a rapid increase within the initial 24 h, after which it stabilized at ~75% throughout the duration of the stability test. Notably, no indications of deactivation are observed, and the selectivity towards $C_{2+}$OH, methanol, $C_{2+}$H, $CH_4$, and $CO_2$ remains consistent throughout the experiment, underscoring the stability of the Cu/Co-CW-$W_1P_1$ catalyst.

Additionally, as displayed in **Figure 8b**, the selectivity of $C_{2+}OH$ exhibits an initial increases within 24 h, followed by stability for the remainder of the stability test, while the distribution of methanol exhibits the opposite trend. This dynamic behavior underscores the importance of stability evaluations over prolonged durations, shedding light on the evolving nature of the catalytic process.

Compared with reported CuCo-based catalysts for the direct conversion of syngas to higher alcohols, the Cu/Co-CW-$W_1P_1$ catalyst displayed exceptional space time yield toward alcohols, indicating superior performance in both selectivity and activity. These findings underscore the potential of the Cu/Co-CW-$W_1P_1$ catalyst for industrial applications in the synthesis of higher alcohols from syngas.

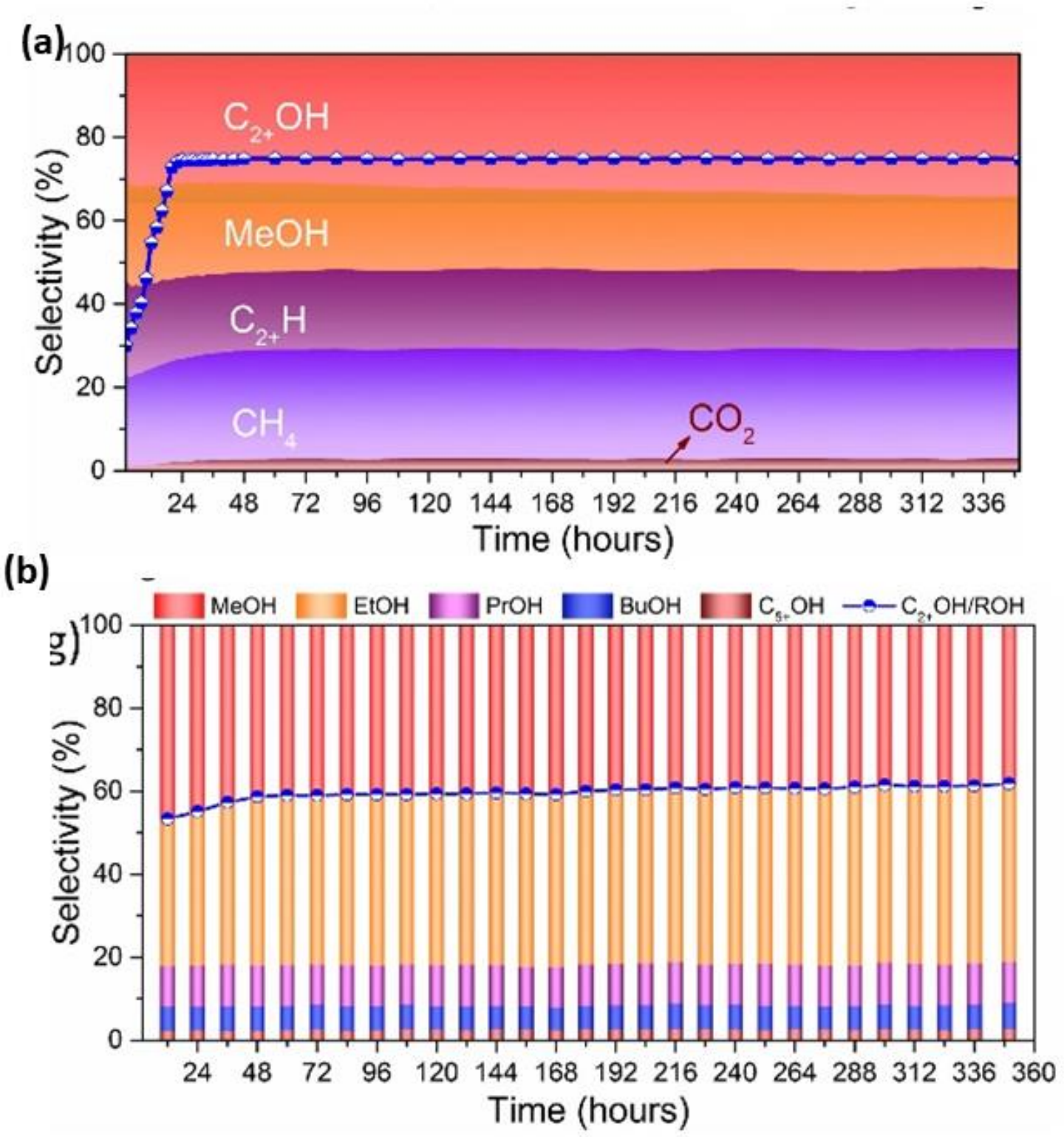

**Figure 8. The catalytic performance of the Cu/Co-CW-$W_1P_1$ catalyst.** (a) A stability test, and (b) the product distribution during 350 h.

## Conclusion

The delicate integration of experimental and theoretical methodologies has enabled us to pinpoint active sites within the low-cost and highly effective 3D carbon wood catalysts. Among these catalysts, the Cu/Co-CW-$W_1P_1$ catalyst emerged as exceptional, showcasing superior selectivity, stability, and activity in the hydrogeneration of CO to higher alcohols, notably $CH_3OH$ and $C_2H_5OH$. The exceptional performance of the Cu/Co-CW-$W_1P_1$ catalyst can be attributed to its unique properties: (i) Open channels with diameters ranging from 5 to 60 μm, facilitating efficient gas flow within the CW with minimal resistance to the wall surface; (ii) Anchored Cu/Co nanoparticles on the CW wall are bidisperse, with two distinct sizes: 15-30 nm and 150-300 nm. The smaller nanoparticles mainly contain $Co_2C$, while the large ones encompass CuO, Cu metal, and a minor quantity of Co metal; (iii) Cu/Co nanoparticles are coated with a thin layer of graphitic carbon, providing protection against coking or sintering.

DFT analysis corroborated the potential for syngas conversion to $CH_3OH$ on the $Co_2C(020)$, while poisoning occurred on $Co_2C(101)$. Moreover, $CH_3O$ dissociation on $Co_2C$ surfaces was determined to be infeasible. Notably, Cu-based surfaces exhibited weaker binding of $CH_3OH$ than Co-based surfaces, suggesting easier desorption. Based on the selectivity descriptor of CO binding energy, CO hydrogenation towards alcohols was found to be achievable across all studied models. This study showcases an efficient, simple, and cost-effective approach utilizing natural wood to prepare carbon wood with exceptional catalytic performance. This advancement sets the stage for the conversion of coal or biomass into higher alcohols, thereby facilitating the further commercialization of this process.

## Methodology

## <u>Experimental section</u>

*Materials*

Balsa wood was bought from JUFASHION Co. Ltd., China, and subsequently cut to dimensions of 3×20×50 mm$^3$. Cobaltous nitrate ($Co(NO_3)_2$•$6H_2O$, Aladdin), cupric nitrate ($Cu(NO_3)_2$•$3H_2O$, Sigma-Aldrich), 1,2-propanediol (Sigma-Aldrich), and deionized water were used in the preparation of all solutions.

*Fabrication of the carbonized wood catalyst*

The balsa wood (3 mm×20 mm×50 mm) was dried at 80 °C for 24 h to remove the moisture. Subsequently, solutions of 0.1 M $Co(NO_3)_2$•$6H_2O$ and 0.1 M $Cu(NO_3)_2$•$3H_2O$ were added into the deionized water or a mixture of water and 1,2-propanediol. The dried wood was soaked into the aforementioned solution for 24 h, followed by drying at 60 °C for an additional 24 h. The treated wood was then transferred to a tubular furnace (OTF-1200X), initially stabilized at 260 °C for 3 h, and subsequently carbonized at 800 °C with a heating rate of 5 °C min$^{-1}$ under $N_2$ flow for 6 h. Finally, the carbon wood was obtained.

*Catalyst characterization:*

The crystal structure analysis of both natural wood and carbon wood were carried out using X-ray diffraction (XRD, Rigaku Ultima IV, Japan), with a scanning range of 10° ~ 90° (2$\theta$) and a scanning speed of 2 s per degree.

The microstructure of the natural wood and carbonized wood was characterized using field emission scanning electron microscopy (FE-SEM, SUPRA 55VP, ZEISS, Germany), operating at 10 kV with a secondary electron pattern, as well as Cs-corrected high-resonlution transmission microscopy (HR-TEM, JEOL, JEM-2200FS) with gold mesh, coupled with an energy-dispersive X-ray (EDX) spectrometer, at an acceleration voltage of 200 kV.

X-ray photoelectron spectroscopy (XPS) was used to analyze the change in the valence state of metal ions. This analysis utilized a Thermo ScientificTM K-AlphaTM+ spectrometer equipped with a monochromatic Al K$\alpha$ X-ray source (1486.6 eV) operating at 100 W. Samples were analyzed under vacuum (P < $10^{-8}$ mbar) with a pass energy of 150 eV for survey scans or 50 eV for high-resolution scans. All peaks were calibrated using the C1s peak binding energy at 284.8 eV for adventitious carbon.

The porosity of CW, Cu/Co-CW-W, Cu/Co-CW-$W_1P_1$, and Cu/Co-CW-$W_1P_2$ was evaluated with the Brunauer Emmett Teller (BET) and Barret-Joyner-Halenda (BJH) method. Prior to analysis, all samples were dried for 6 h in a vacuum over at 70 °C.

Thermogravimetric analysis (TGA) of the catalysts was performed under an air flow using a simultaneous thermal analyzer (NETZSCH, STA-449-F5).

Raman spectrum of the catalysts was acquired using a Thermo Scientific Dxi2 Raman spectrometer with laser excitation at a wavelength detector of 532 nm.

*Product analysis*

The catalytic performance of the Cu/Co-CW-W, Cu/Co-CW-$W_1P_1$, and Cu/Co-CW-$W_1P_2$ catalyst for higher alcohol synthesis was evaluated using a Micro-Activity Effi (Micromertics Instrument Ltd, USA), as shown in our previous work.[35] The obtained Cu/Co carbon wood was loaded into the reaction tube. The reaction condition was set to 3 MPa, 270 °C, $H_2$/CO = 2/1, and GHSV = 9600 mL $g_{cat}^{-1}$ $h^{-1}$. (see **Table S1** for details).

*Performance measurements:*
The catalytic performance of the CuCo@SNTs-c catalyst for higher alcohol synthesis from syngas was investigated using a Micro-Activity Effi from Micromeritics Instrument Ltd, USA, equipped with a tubular fixed-bed reactor. The catalyst was loaded into the middle of the reaction tube. The evaluation was conducted under the condition of 270 °C, 3.0 MPa, syngas (60 vol % $H_2$, 30 vol % CO, and 10 vol % $N_2$), with a GHSV= 9600 mL $g_{cat}^{-1}$ $h^{-1}$. $N_2$ was used as an internal standard. The tail gas containing CO, $H_2$, $N_2$, $CO_2$, $CH_4$, and $C_2H_6$ was analyzed using an Agilent GC 7890B gas chromatograph equipped with a thermal conductivity detector using TDX-1 and Hayesep Q packed column with argon (Ar) as the carrier gas. Liquid products collected from the high-precision gas-liquid separator were analyzed offline using an Agilent GC 7890B with an HP-5 capillary column and a hydrogen flame ionization detector, employing 2-butanol as the internal standard, with Ar as the carrier gas.

CO conversion ($X_{CO}$), product selectivity on carbon basis ($S_i$), space-time yield (STY), and Anderson-Schulz-Flory (ASF) distribution of higher alcohols were calculated using the following equations.

$$X_{CO} = \frac{n_{CO,in} - n_{CO,out}}{n_{CO,in}} \times 100\% \qquad (1)$$

$$S_{CO_2} = \frac{n_{CO_2}}{n_{CO,in} - n_{CO,out}} \times 100\%, \qquad (2)$$

$$S_i = \frac{N_i \times n_i}{\sum(N_i \times n_i)} \times 100\%, \qquad (3)$$

$$STY = \frac{weight\ of\ alcohols\ product\ (g)}{weight\ of\ catalyst\ (g) \times h}, \qquad (4)$$

$$\frac{W_n}{n} = (1-\alpha)^2 a^{n-1} \rightarrow ln\frac{V_n}{n} = nln\alpha + c, \qquad (5)$$

where $n_{CO,in}$ and $n_{CO,out}$ were the moles of CO at the inlet and outlet, respectively. The $N_i$ and $n_i$ denoted the moles and carbon number of alcohol product, respectively.

## Computational section

All simulations were performed using spin-polarized DFT within the framework of the Vienna Ab initio Simulation Package.[36, 37] The Generalized Gradient Approximation (GGA) of the Perdew-Burke-Ernzerhof (PBE) exchange-correlation functional with the D2 dispersion correction (PBE-D2) was the functional of choice.[38, 39] Core electrons were described by projector augmented wave,[37, 39, 40], while valence electrons were expanded by plane wave basis sets with a kinetic energy cut-off of 400 eV. The catalyst surfaces considered included $Co_2C$ (101, 020), CuO (111), and Cu(100), each comprising at least four layers, with the top two layers fully relaxed and the remaining layers constrained to bulk distances. Structural relaxation was performed with a force convergence threshold of 0.05 eV/Å. The Brillouin zone was sampled using a Γ-centered (4 × 5 × 1), (5 × 5 × 1), (3 × 4 × 1), and (5 × 5 × 1) k-points mesh for $Co_2C(101)$, $Co_2C(020)$, CuO(111), and Cu(100) respectively, generated using the Monkhorst−Pack method.[41] The vacuum between the slabs was set to 15 Å in the z direction.

For all studied models, the Gibbs energies for the thermal path of pertinent intermediates were determined using ΔG = ΔH + ΔZPE − TΔS. Gas-phase molecules were used as references and were calculated with the PBE functional. The zero-point and entropic contributions are provided in **Tables S10-S14**. Additional details regarding thermodynamic corrections can be found in the Supporting Information. All structural data are available for retrieval from the Zenodo repository via the following link:(https://doi.org/10.5281/zenodo.8239714).

## Acknowledgments

G.Y. thank the China Scholarship Council for its financial support. P.P. is thankful for the financial support from the Academy of Finland under the C1 Value program (No. 329483). M.A.C., X.C., B.P. acknowledge the financial support from the Academy of Finland Fellow grant, No. 330488, No. 335571, and No. 321443, respectively. Computational resources for this project were obtained from the CSC—IT Center for Science and Aalto University's Science-IT project. G.Y., G.C., and Y.L. acknowledge the financial support from the National Key R&D Program of China (No. 2021YFC2101604), National Natural Science Foundation of China (No. 22278339), and Innovative Funds Plan of Henan University of Technology (No. 2022ZKCJ09).

## Supporting information

The supporting information includes Figure S1 to Figure S27, Table S1 to Table S4, and the computational details (Tables S5-S14), *etc*. SEM images and SEM-EDS of the samples, TEM images and the TEM-EDX of the samples, and XRD, XPS, Raman, TG, and BET spectra of the samples.

# Supporting Information

## Syngas conversion to higher alcohols via wood-framed Cu/Co-carbon catalyst

*Guihua Yan[1,2,3], Paulina Peršlja[4], Gaofeng Chen[1,3], Jiahui Kang[2], Yongde Liu[1], Miguel A. Caro[4], Xi Chen[2,5], Xianhai Zeng[3]*, Bo Peng[2,6]**

[1] College of Environmental Engineering, Henan University of Technology, 450001, China

[2] Department of Applied Physics, Aalto University, FI-00076 Aalto, Finland

[3] College of Energy, Xiamen University, 361102, China

[4] Department of Chemistry and Materials Science, Aalto University, Kemistintie 1, 02150 Espoo, Finland

[5] Present address: The School of Physical Science and Technology, Lanzhou University, 730000, China

[6] Department of Materials Science, Advanced Coatings Research Center of Ministry of Education of China, Fudan University, 200433, China

E-mails: xianhai.zeng@xmu.edu.cn, peng_bo@fudan.edu.cn

Guihua Yan, Paulina Peršlja, and Gaofeng Chen contributed equally.

# 1. Supporting results and discussion

## 1.1. Structural characterization

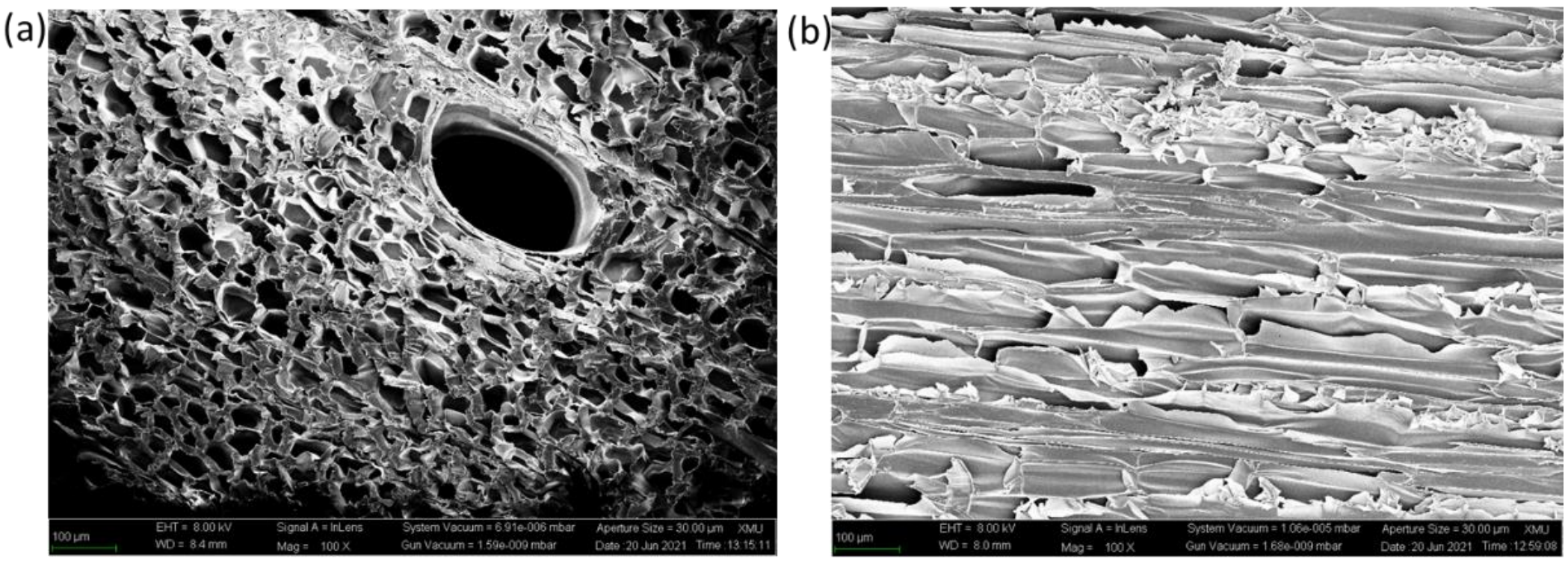

**Figure S1**. SEM images of natural wood from (a) R-view and (b) L-view.

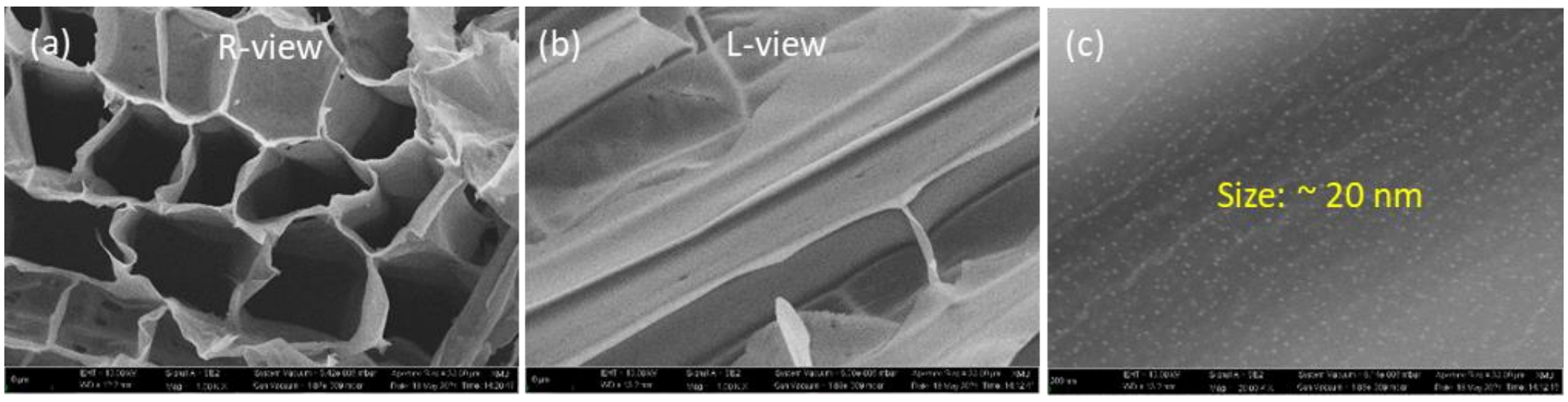

**Figure S2**. SEM images of carbon wood impregnated with deionized water as a solvent.

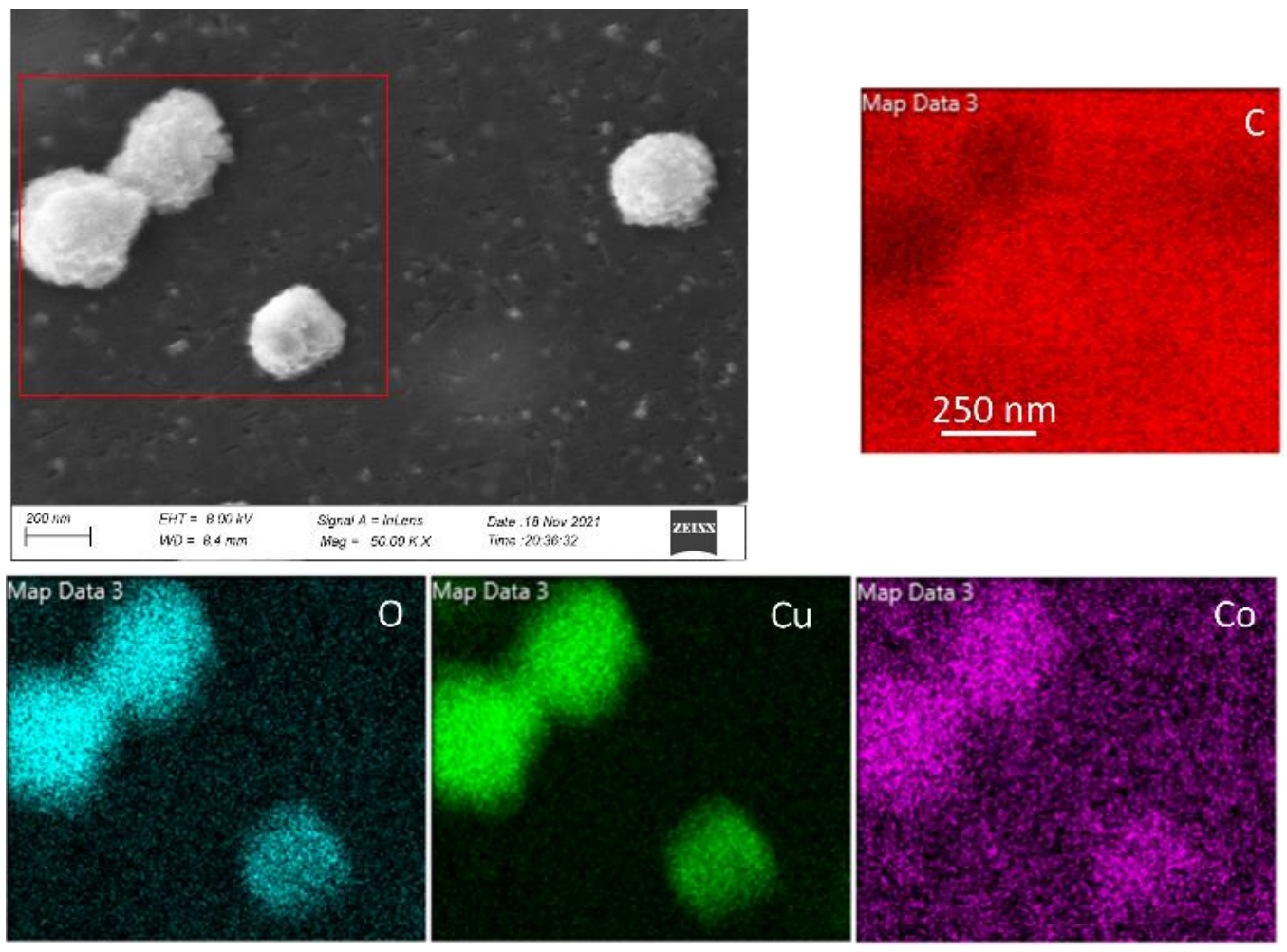

**Map Sum Spectrum**

| Element | Weight % | σ |
|---|---|---|
| C | 67.05 | 3.78 |
| Cu | 15.88 | 0.90 |
| Co | 7.70 | 5.20 |
| O | 7.59 | 0.43 |
| others | 1.79 | 0.15 |

**Figure S3**. The element mapping and contents of Cu/Co carbon wood, corresponding to Figure 2k.

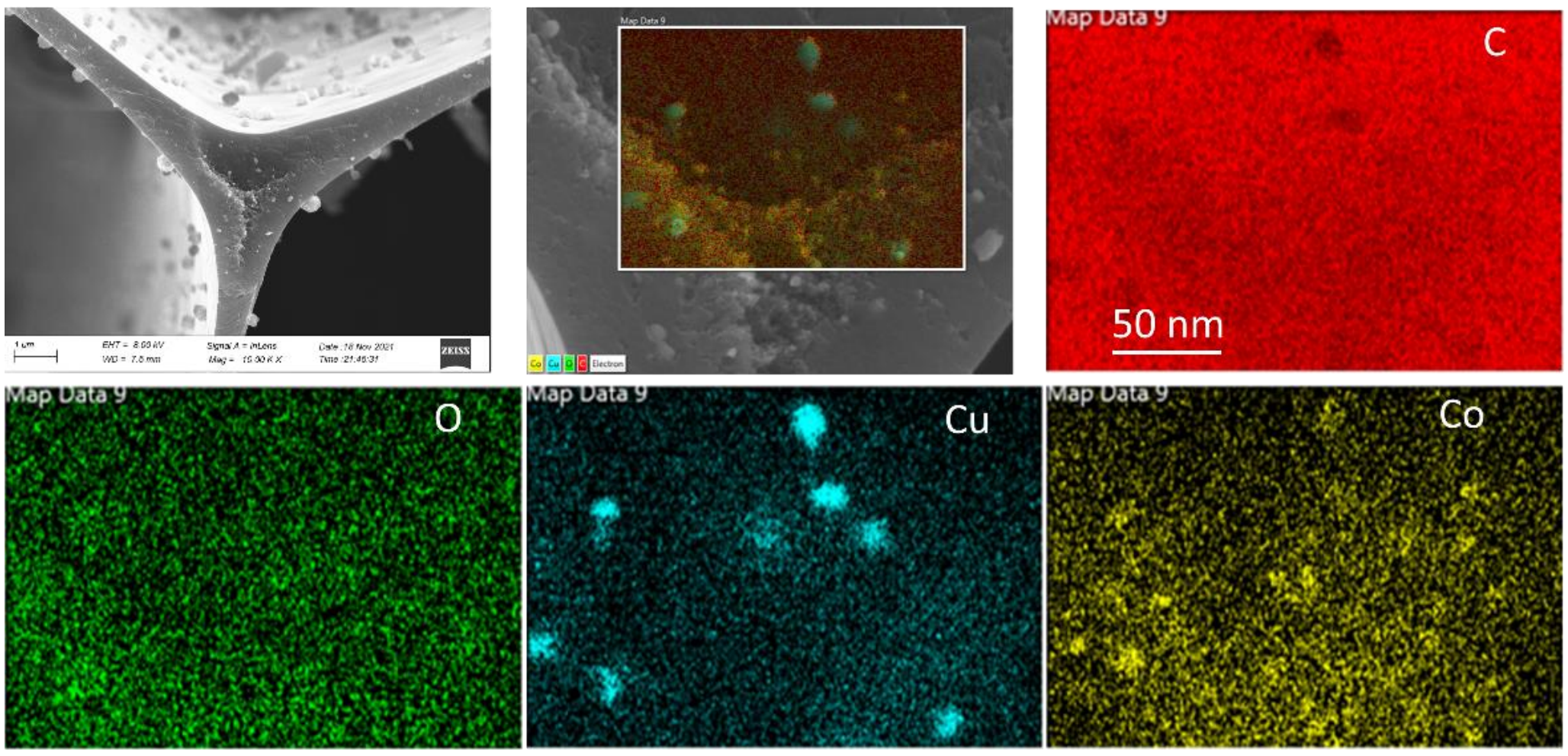

**Map Sum Spectrum**

| Element | Weight % | σ |
|---|---|---|
| C | 85.95 | 4.48 |
| Co | 7.38 | 4.83 |
| Cu | 3.67 | 0.20 |
| O | 2.65 | 0.14 |
| others | 0.36 | 5.36 |

**Figure S4**. The element mapping and contents of the Cu/Co carbon wood, corresponding to Figure 2l.

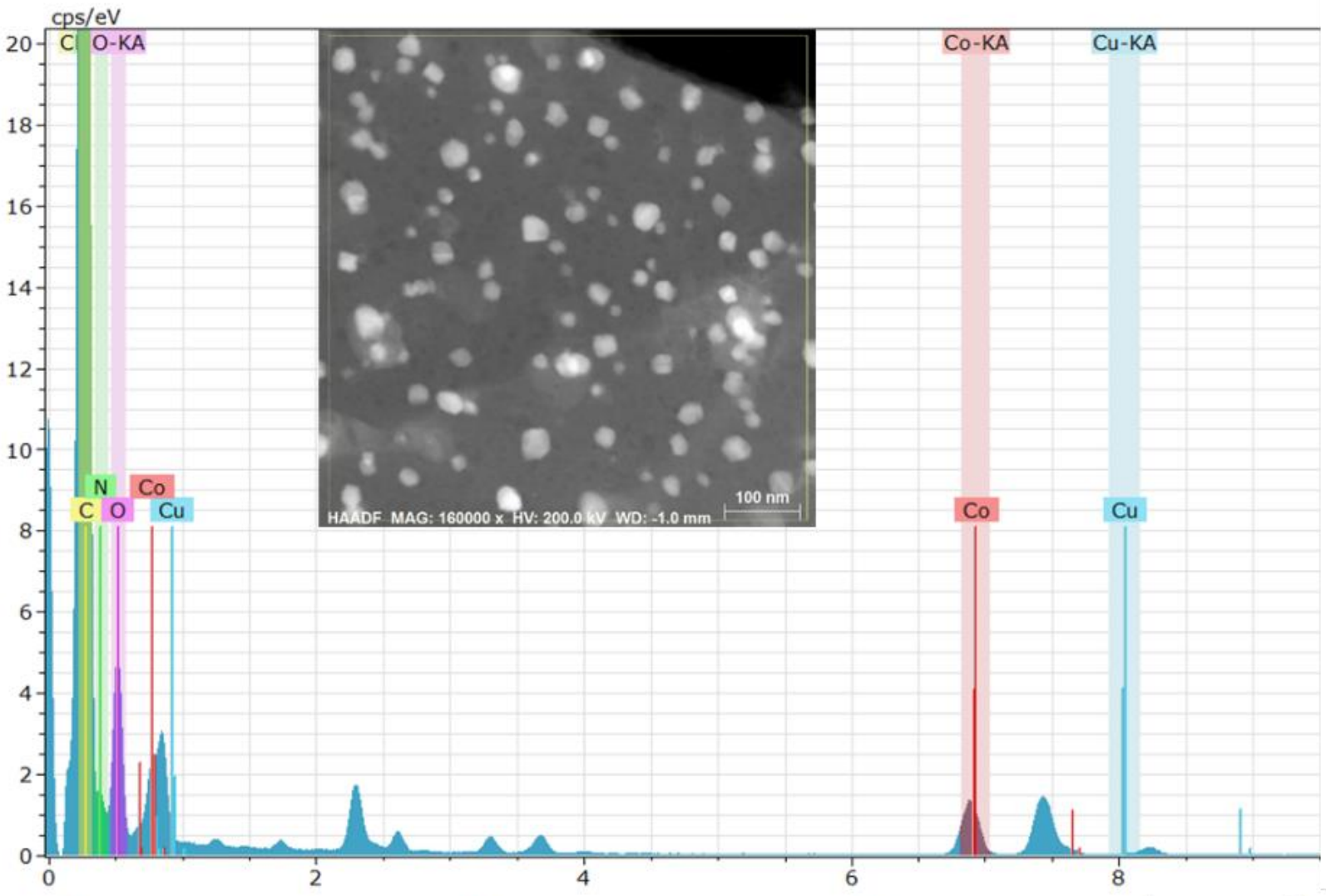

| Element | Series | Net | unn. C [wt.%] | norm. C [wt.%] | Atom. C [at.%] | Error (3 Sigma) [wt.%] |
|---|---|---|---|---|---|---|
| Copper | K-series | 979 | 0.11 | 0.11 | 0.02 | 0.09 |
| Cobalt | K-series | 27809 | 2.39 | 2.39 | 0.50 | 0.29 |
| Carbon | K-series | 661713 | 93.69 | 93.69 | 96.41 | 8.51 |
| Nitrogen | K-series | 13064 | 1.14 | 1.14 | 1.01 | 0.18 |
| Oxygen | K-series | 42240 | 2.67 | 2.67 | 2.07 | 0.32 |
| | | Total: | 100.00 | 100.00 | 100.00 | |

**Figure S5**. Element EDS of Cu/Co NPs in carbon wood from TEM image.

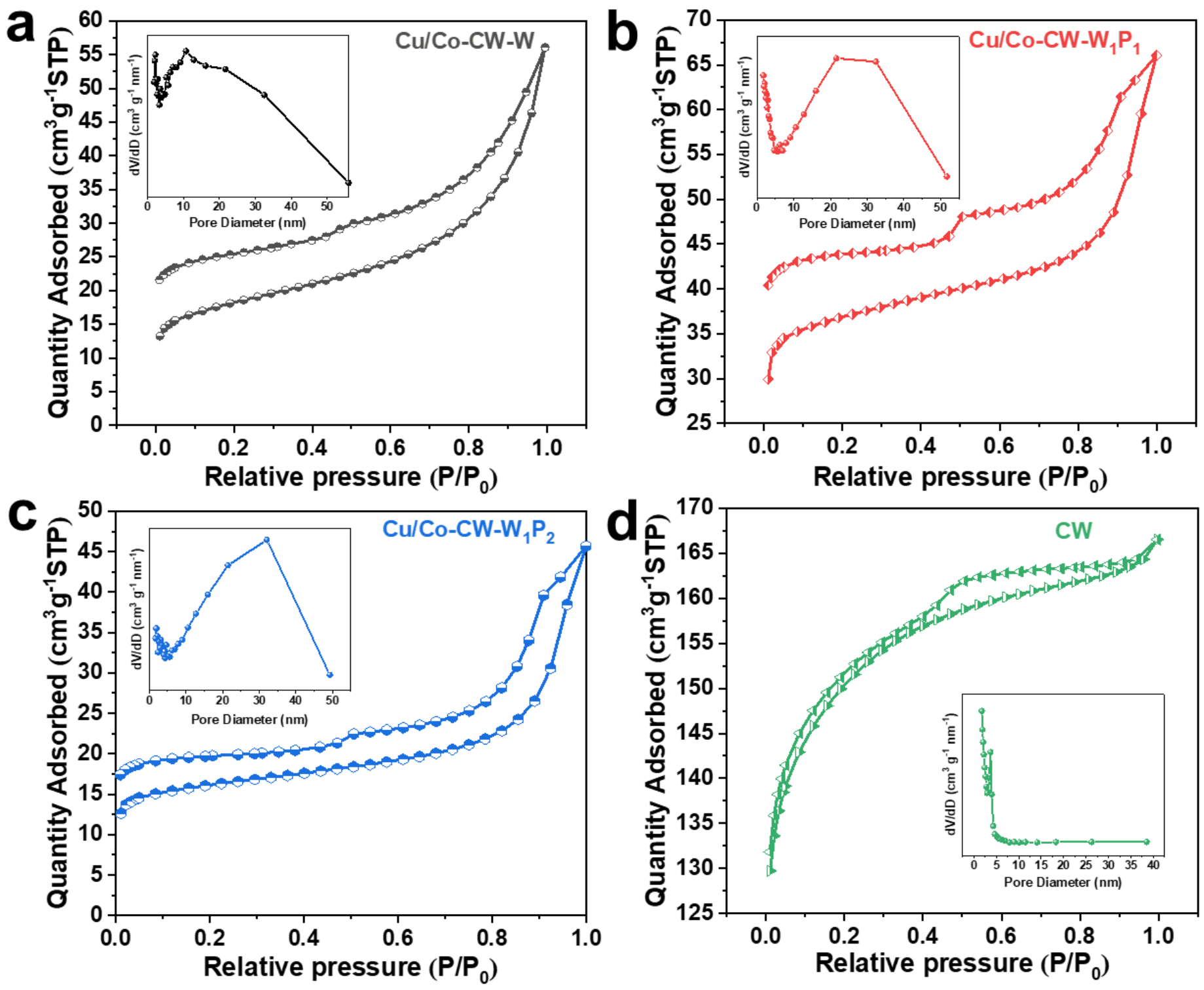

**Figure S6**. BET patterns of the CW, Cu/Co-CW-W, Cu/Co-CW-$W_1P_1$, and Cu/Co-CW-$W_1P_2$.

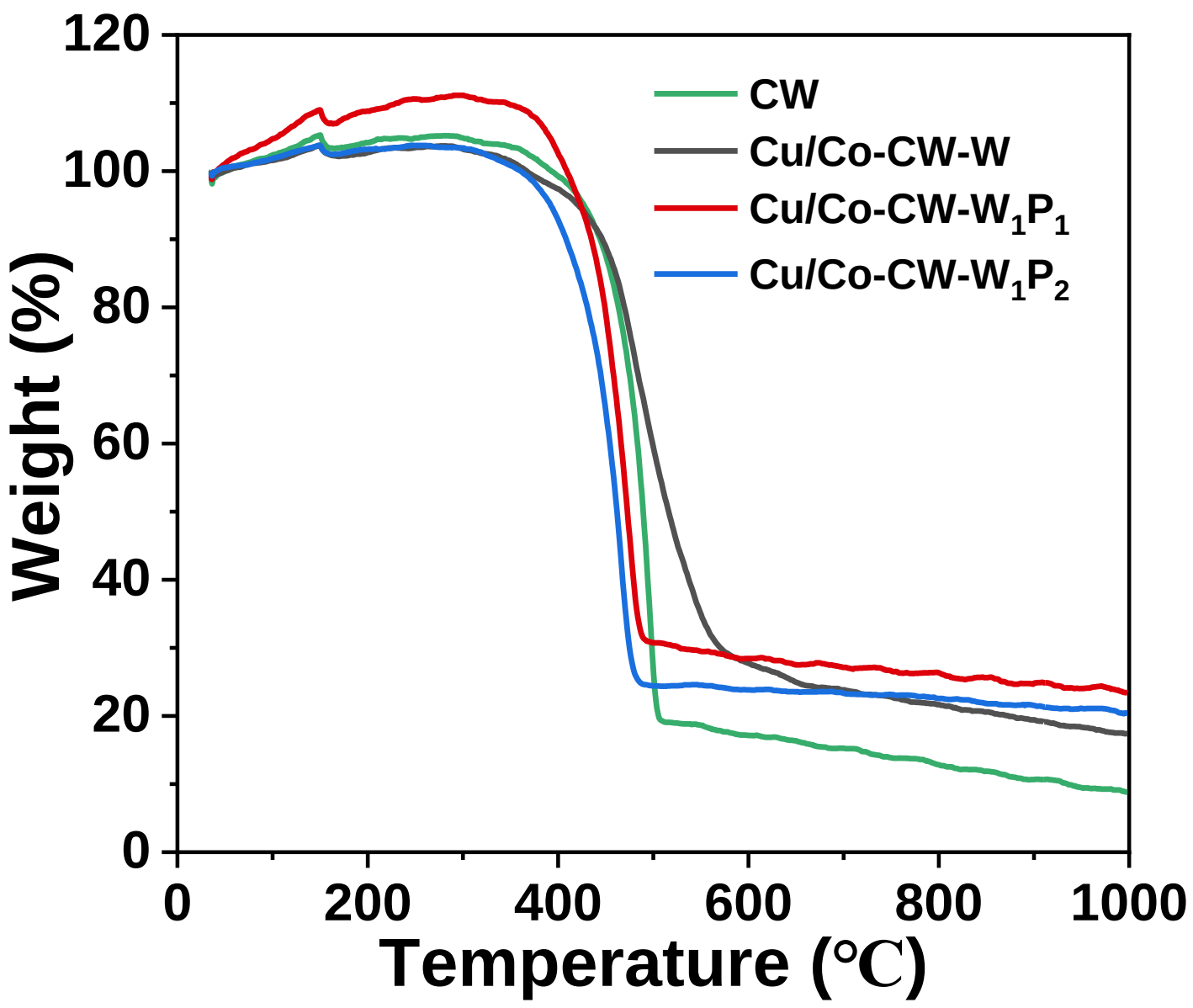

**Figure S7.** TG patterns of the CW, Cu/Co-CW-W, Cu/Co-CW-$W_1P_1$, and Cu/Co-CW-$W_1P_2$.

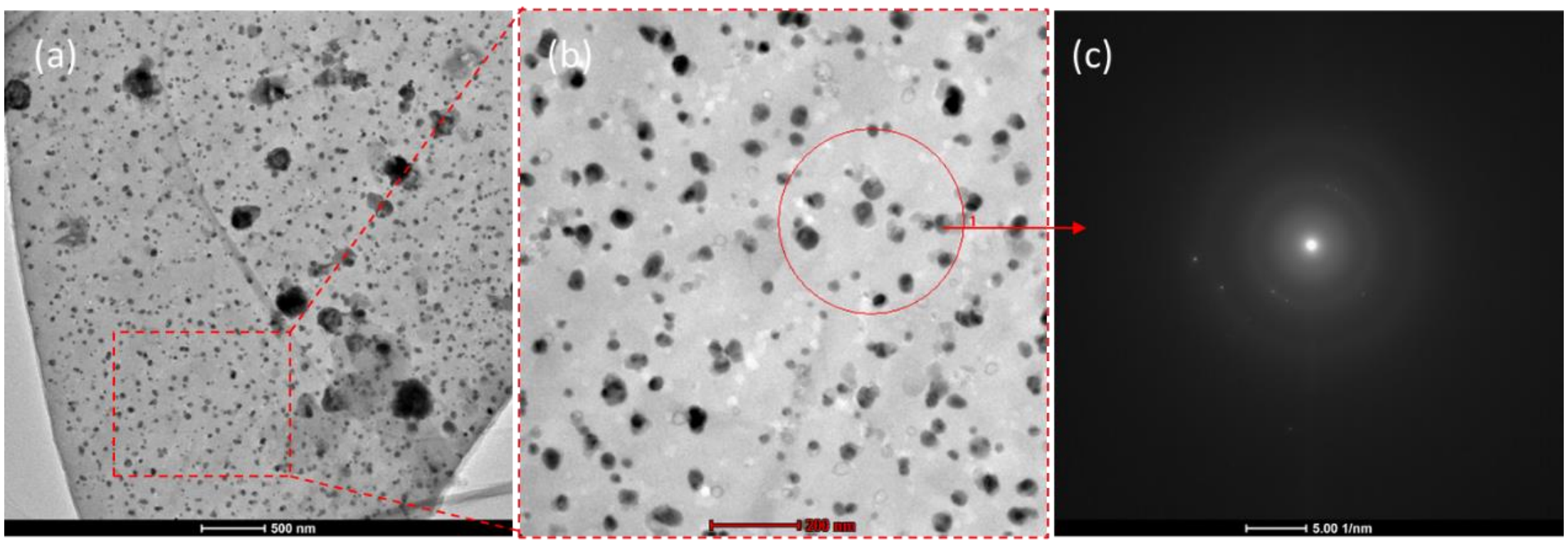

**Figure S8**. (a) Cs-corrected HR-TEM image of Cu/Co-CW-$W_1P_1$ containing bidisperse nanoparticles, and (b) its corresponding enlarged view, and (c) the diffraction pattern of the region highlighted in (b).

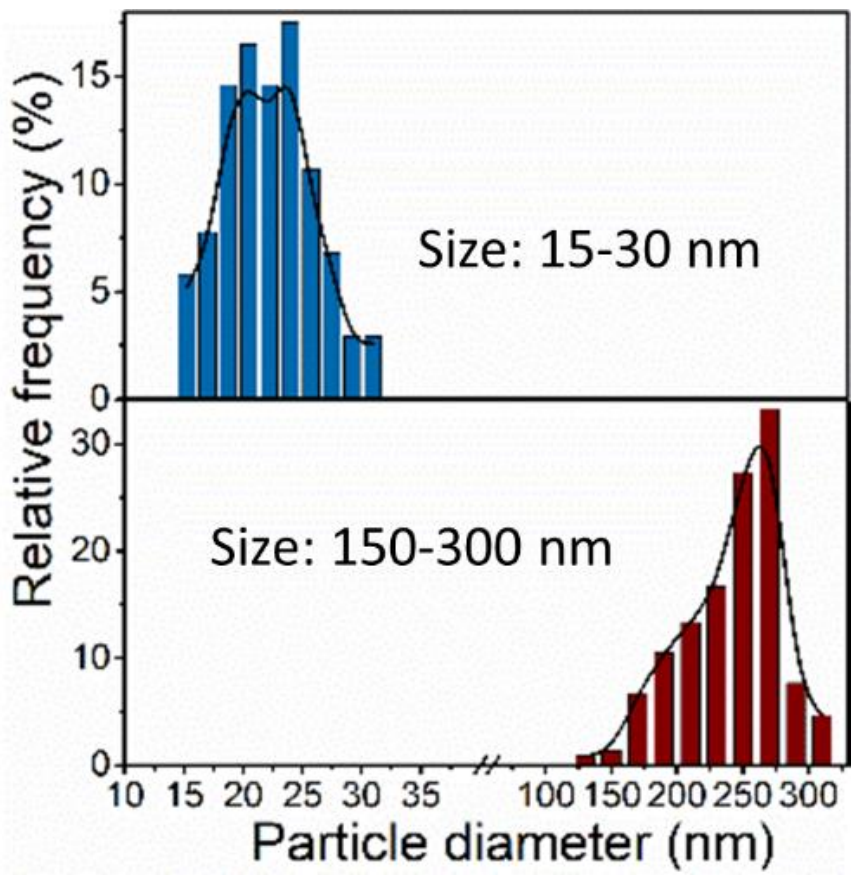

**Figure S9**. Size distribution of bidisperse nanoparticles.

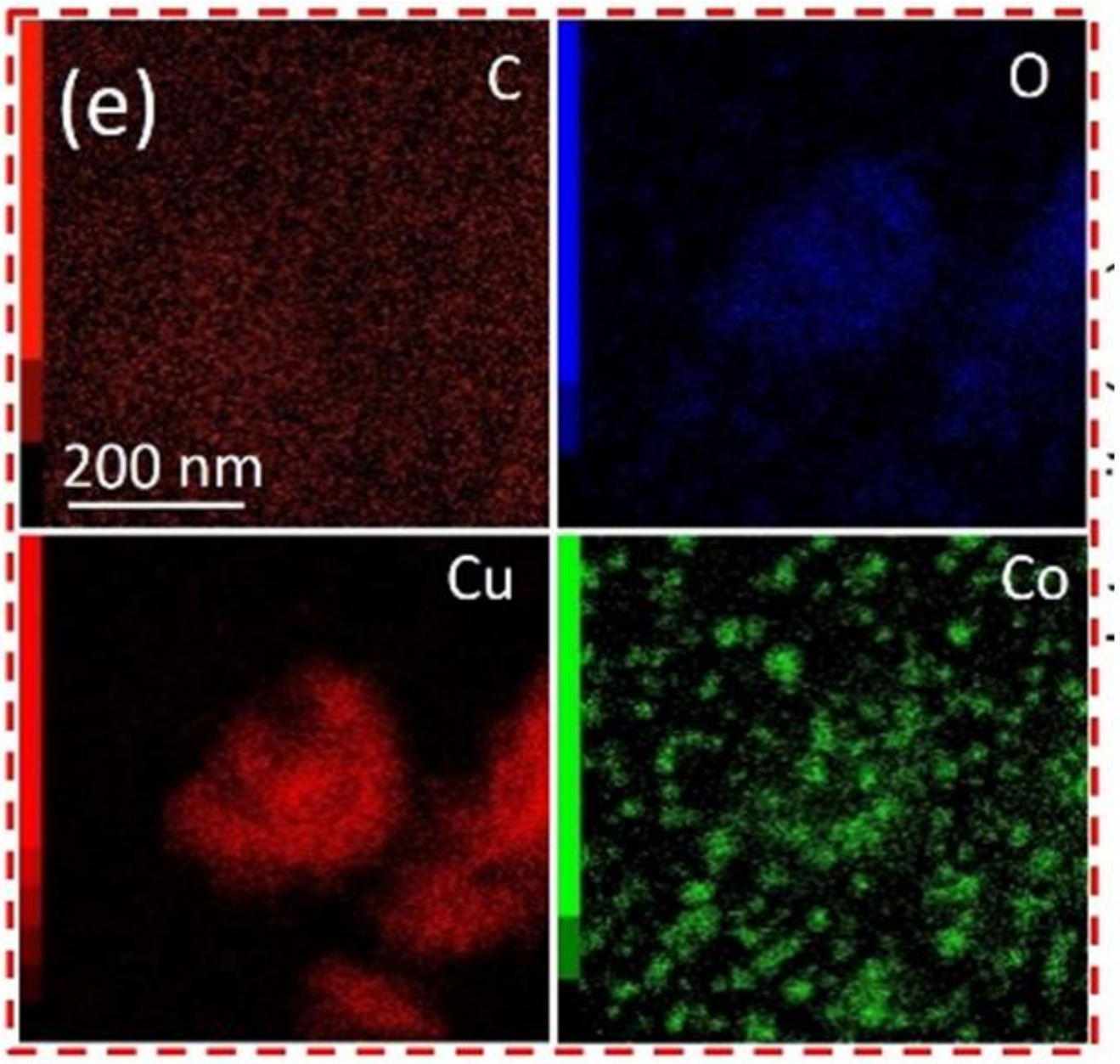

**Figure S10**. EDX elemental analysis of the Cu/Co-CW-$W_1P_1$ sample marked in Figure 3b.

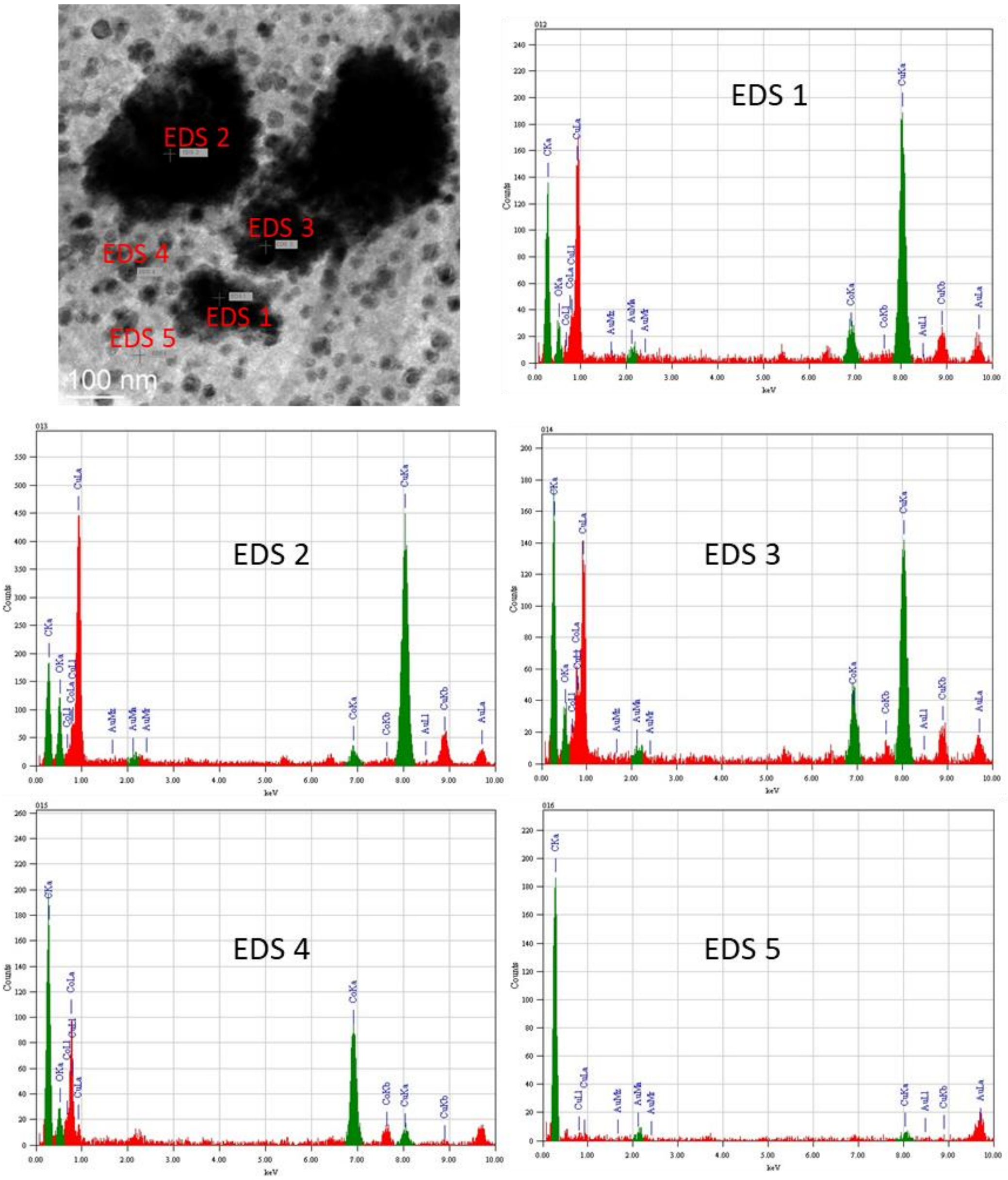

**Figure S11**. Elemental contents of the Cu/Co-CW-$W_1P_1$ sample.

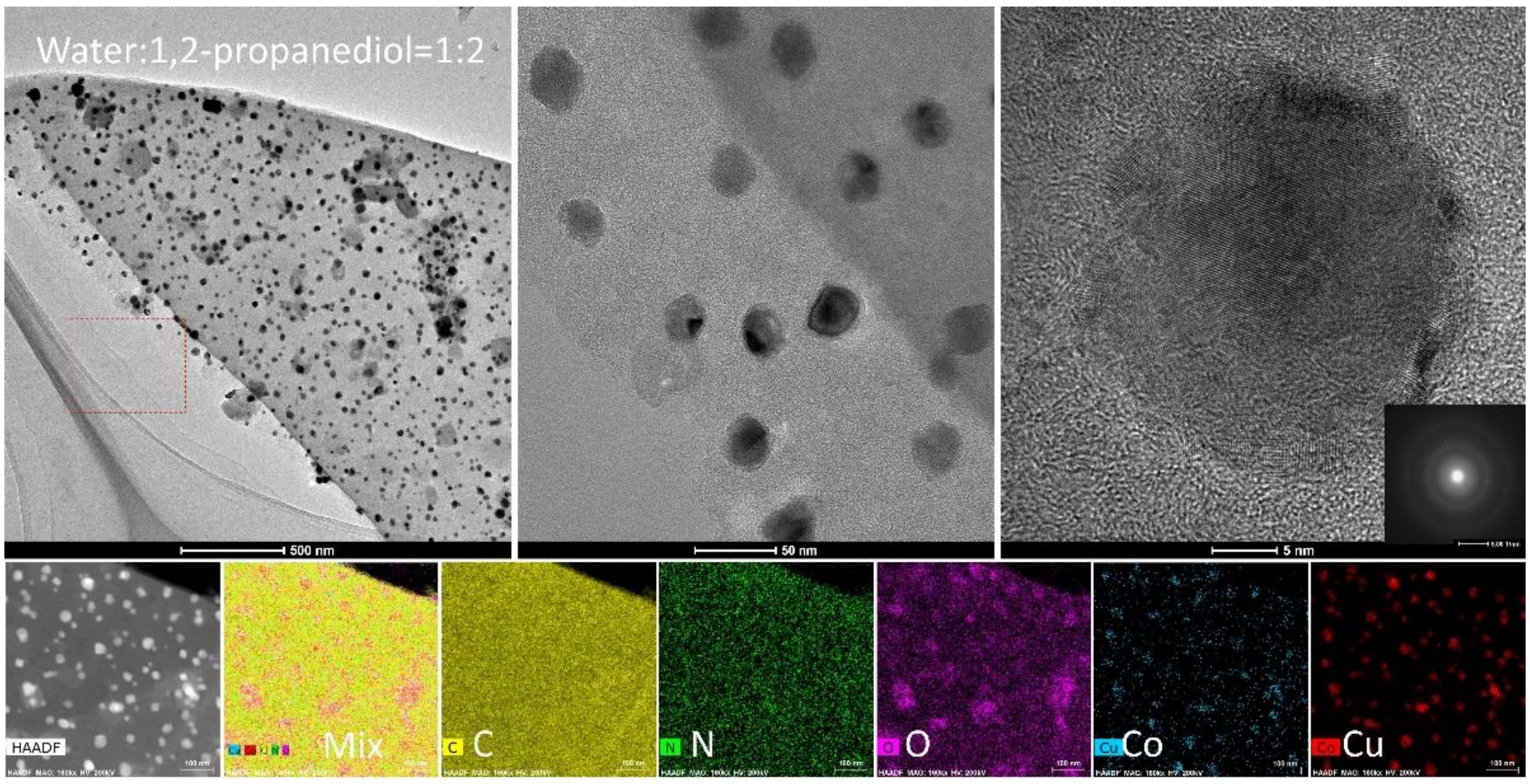

**Figure S12**. Element EDS of Cu/Co NPs (Water:1,2-propanediol=1:2) in carbon wood from TEM image.

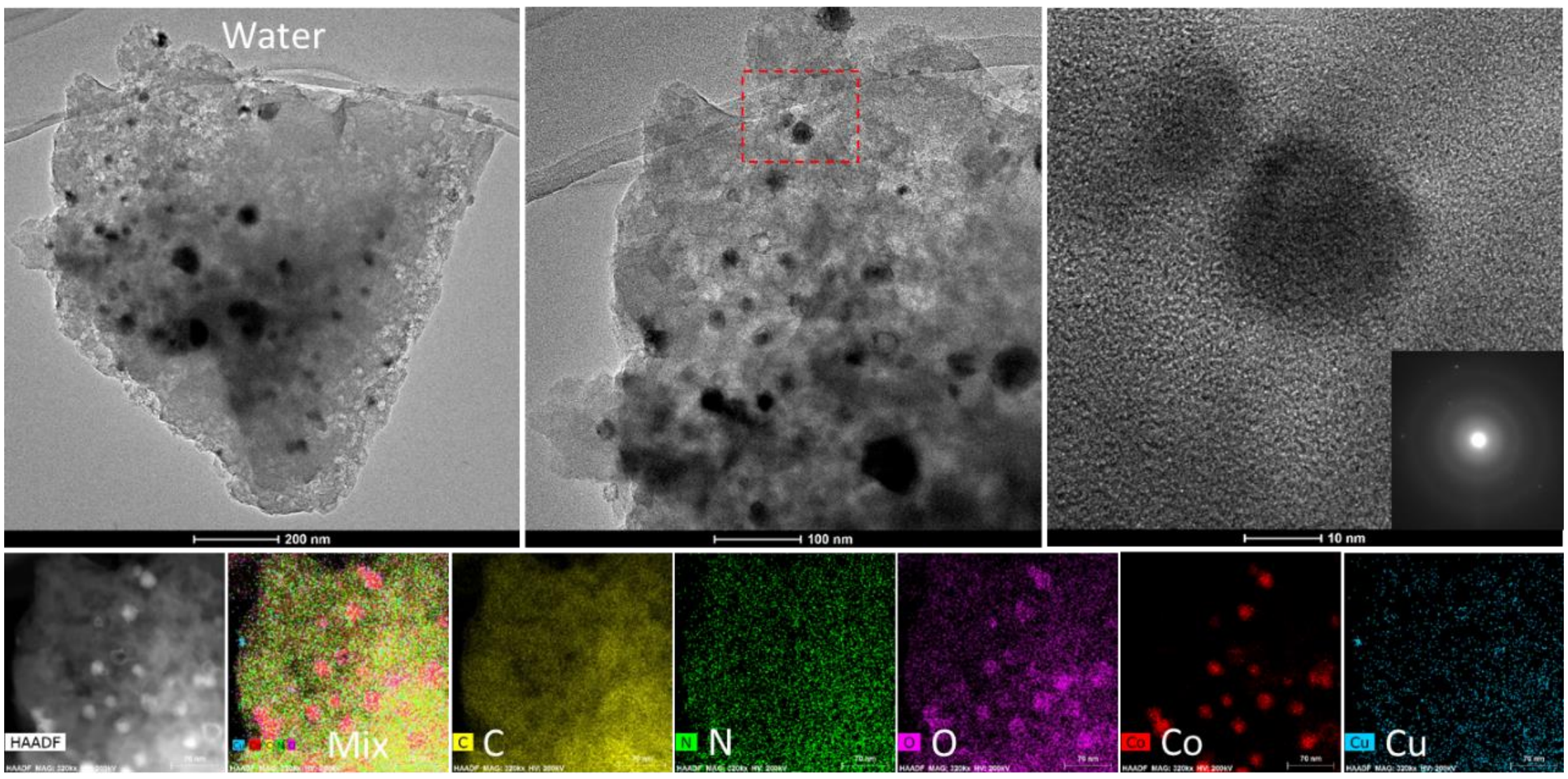

**Figure S13**. Element EDS of Cu/Co NPs (in water) in carbon wood from TEM image.

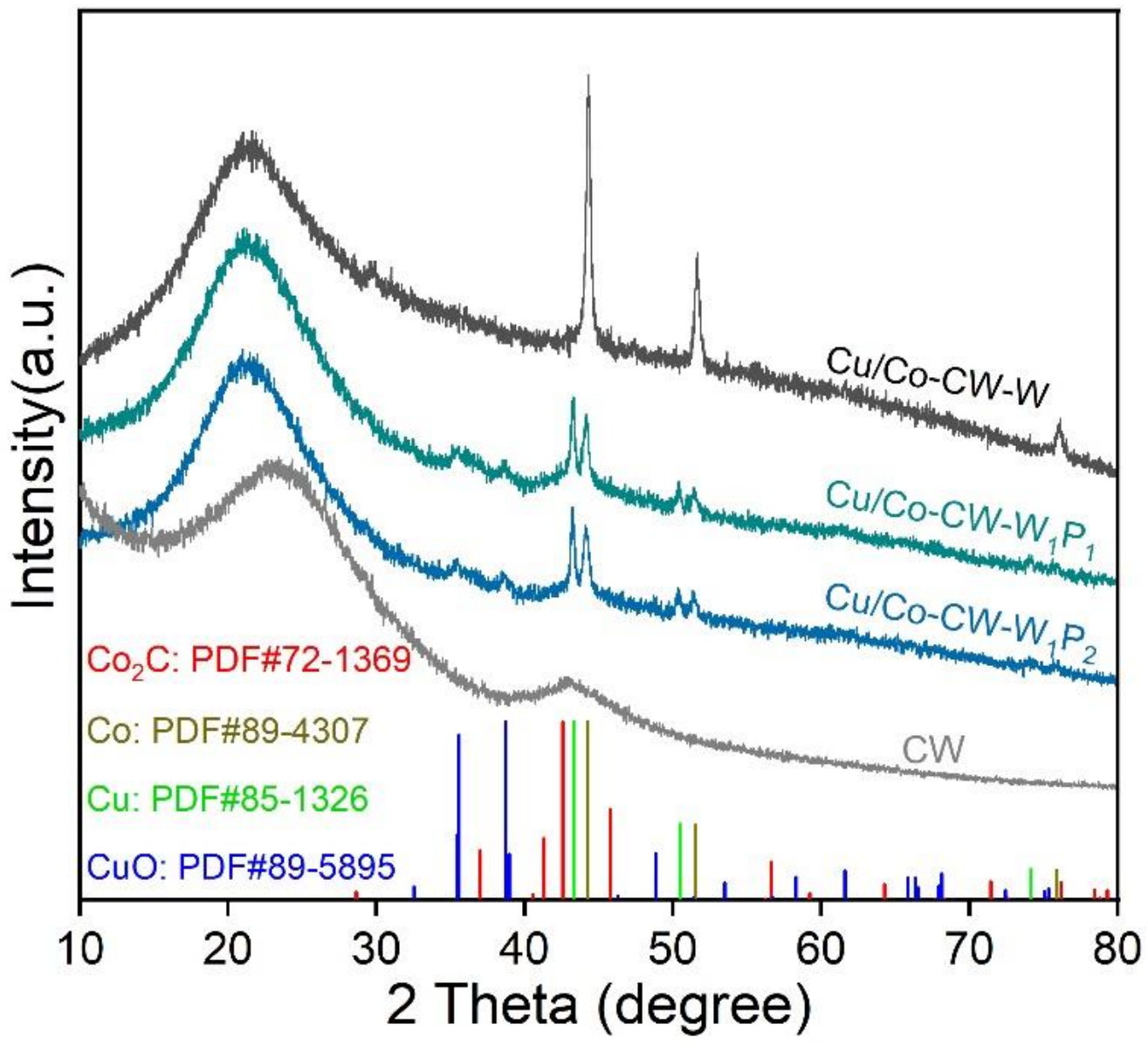

**Figure S14**. The XRD patterns of the CW, Cu/Co-CW-W, Cu/Co-CW-$W_1P_1$, and Cu/Co-CW- $W_1P_2$.

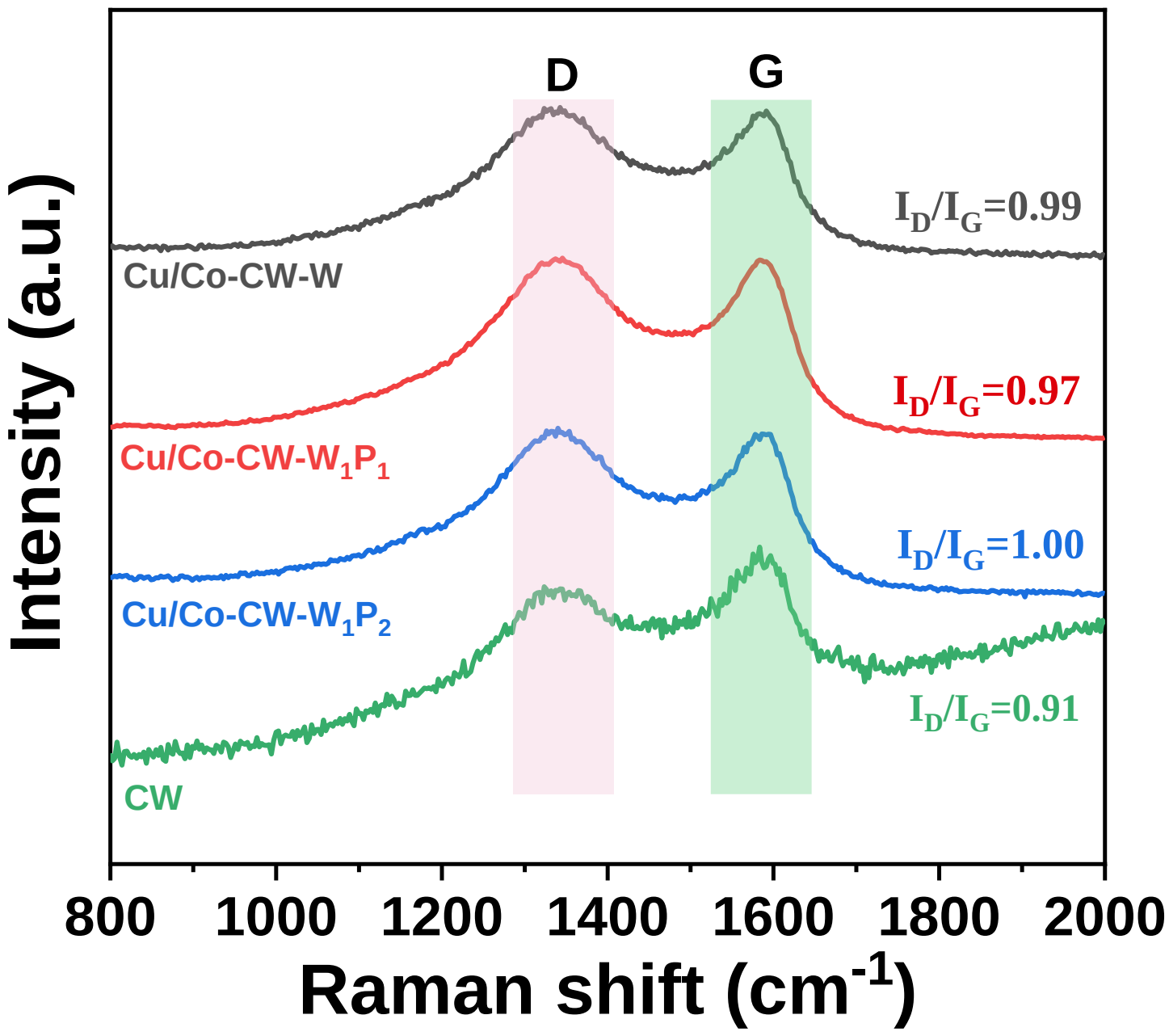

**Figure S15**. Raman spectra of the CW, Cu/Co-CW-W, Cu/Co-CW-$W_1P_1$, and Cu/Co-CW- $W_1P_2$.

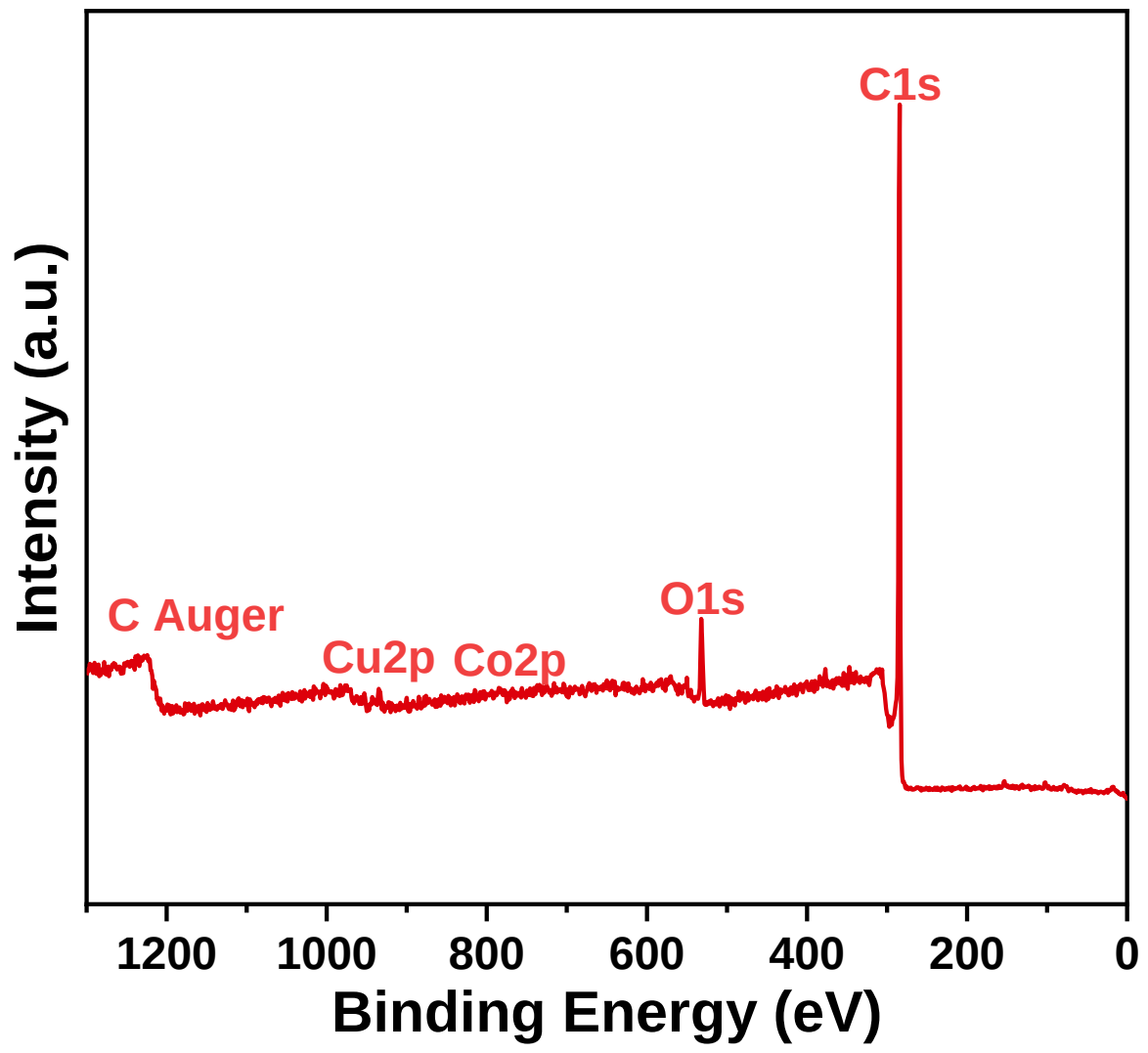

**Figure S16**. The full XPS spectra of Cu/Co-CW-$W_1P_1$.

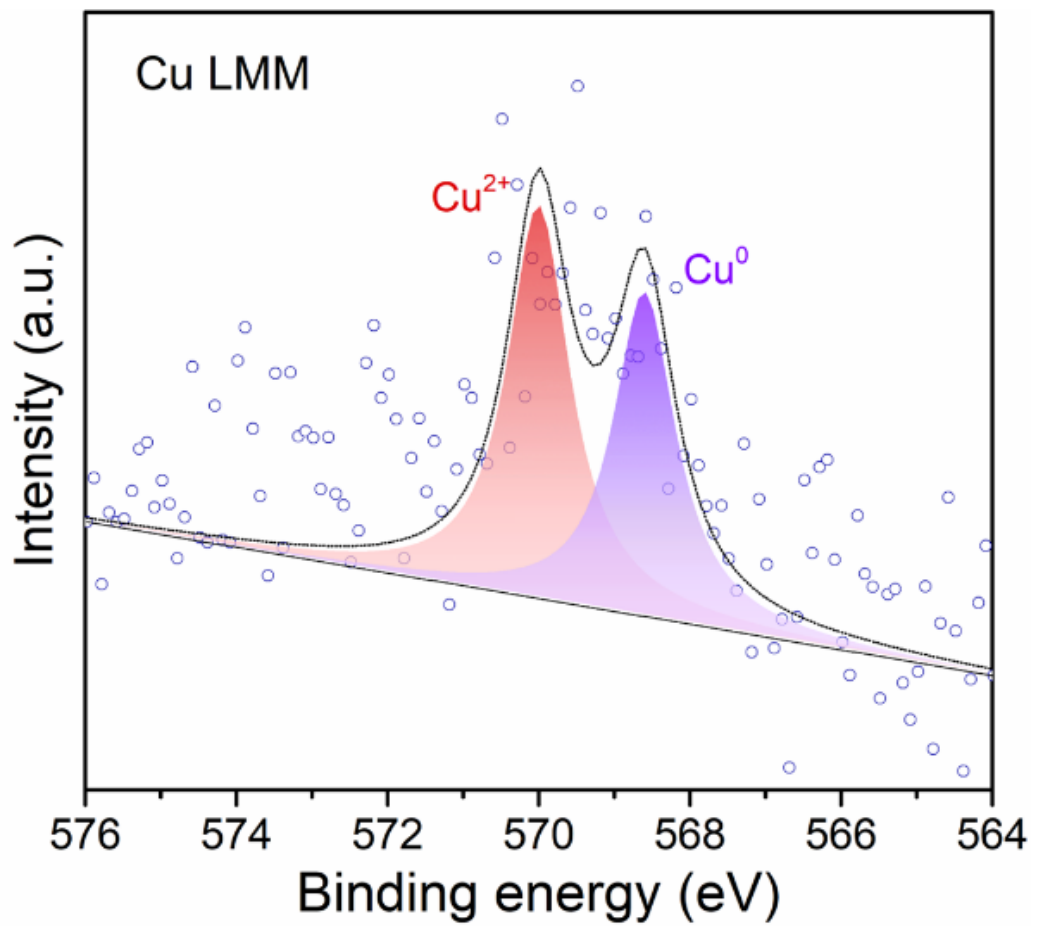

**Figure S17**. The XPS patterns of auger electron peaks of Cu (Cu LMM).

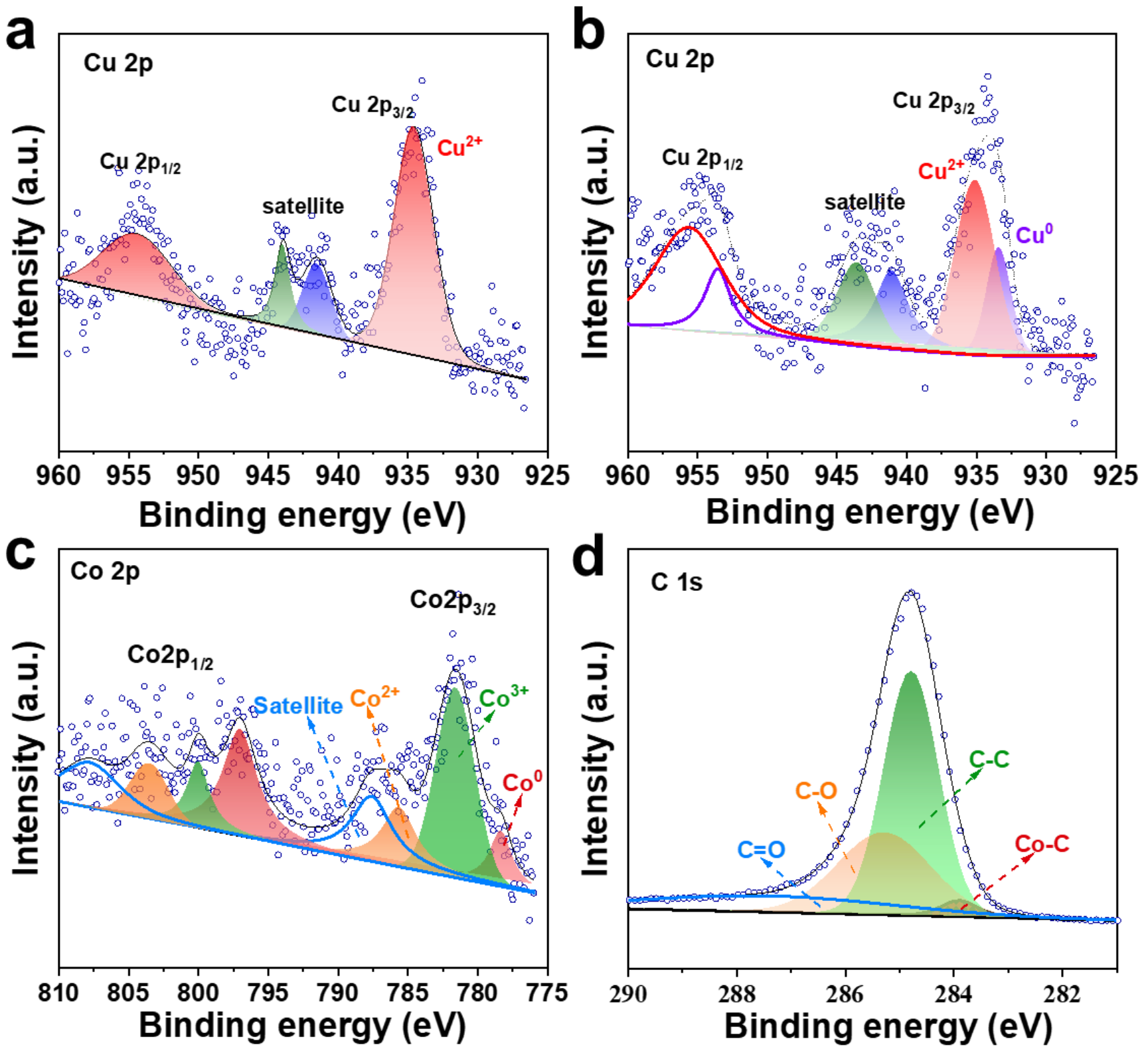

**Figure S18**. The XPS patterns of Cu/Co-CW-$W_1P_2$.

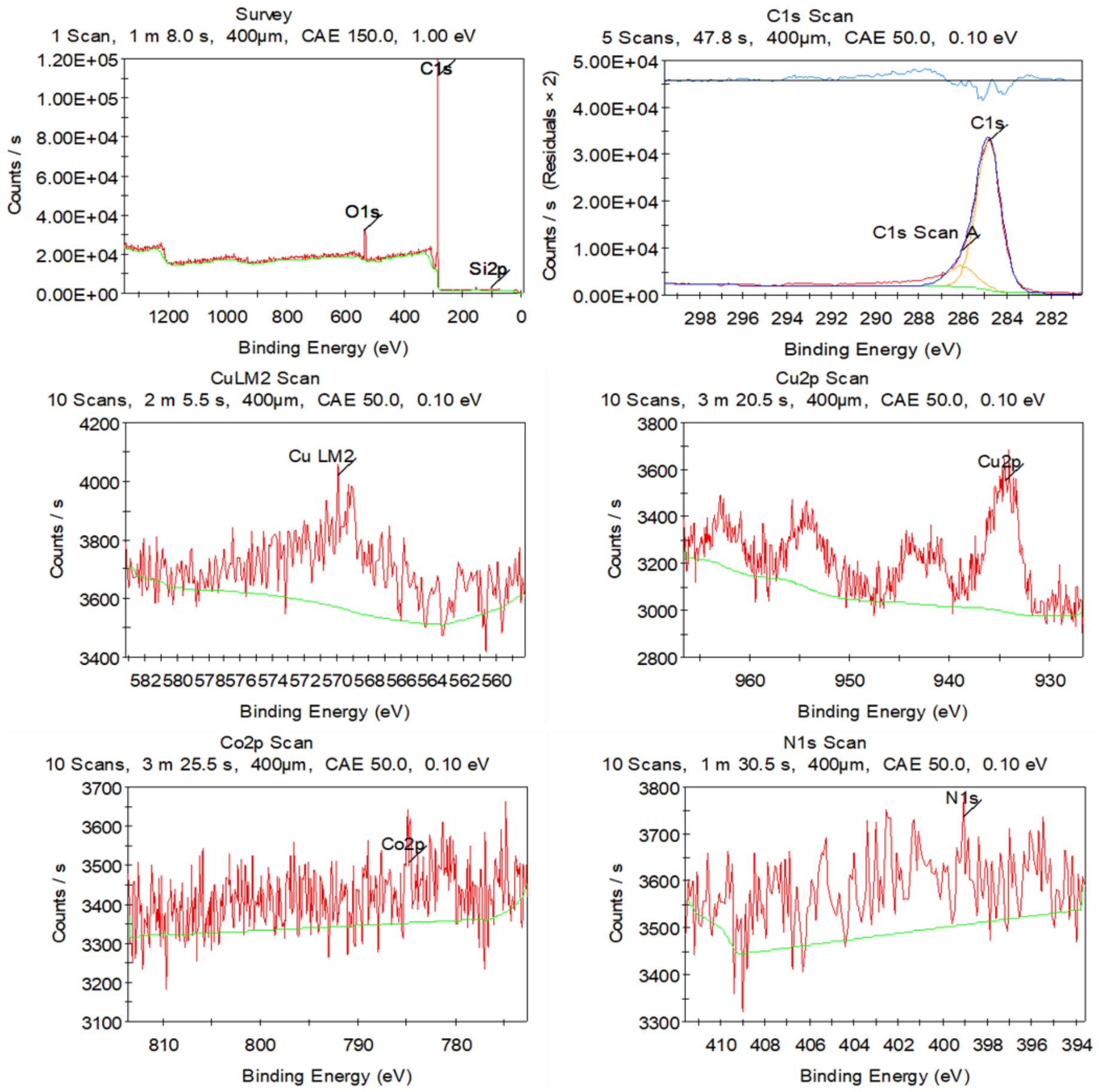

**Figure S19**. The XPS patterns of Cu/Co-CW-W

## 1.2. Catalytic performance

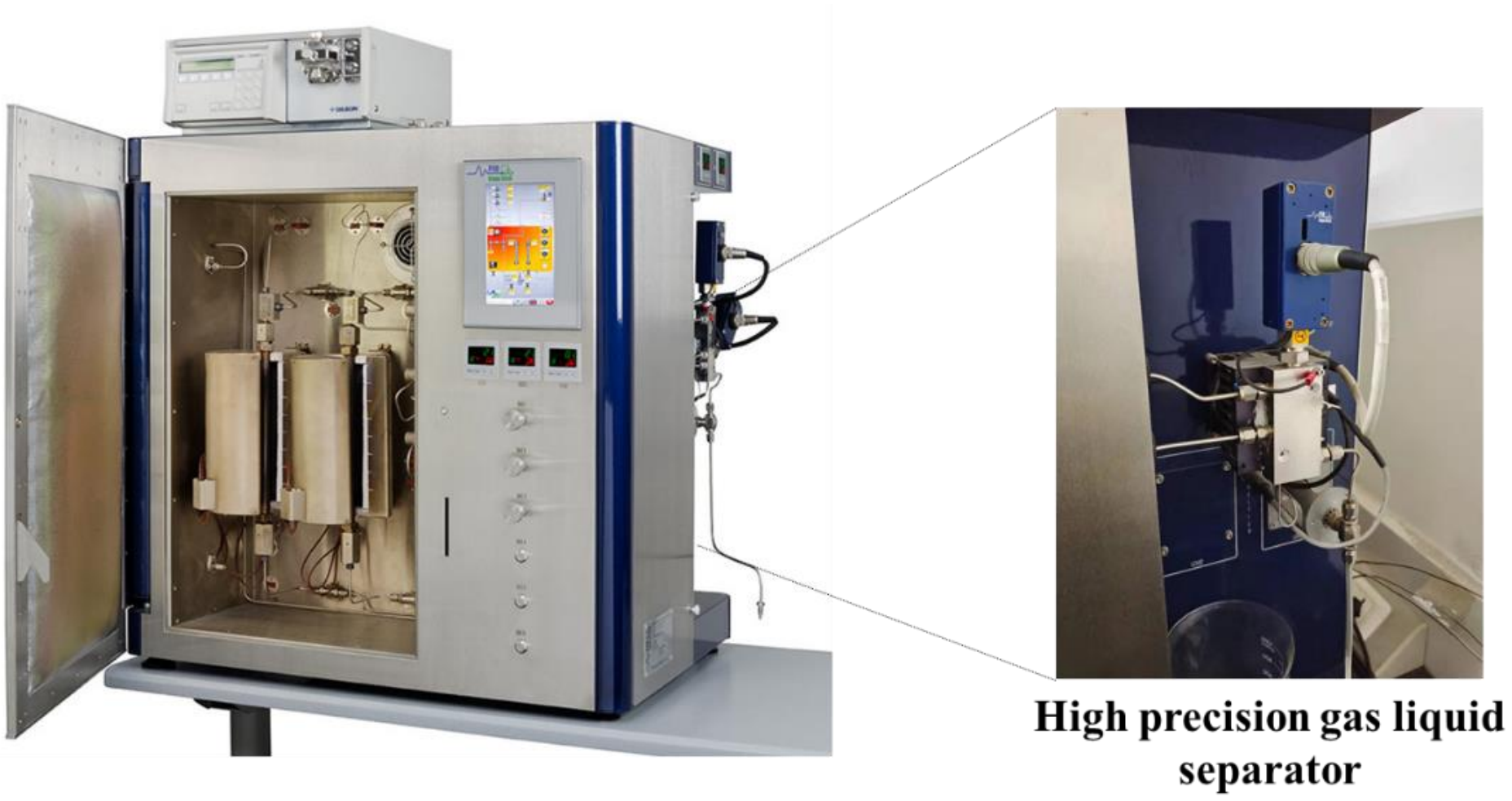

**Figure S20.** Photograph of the Micro-activity Effi with a high precision gas liquid separator.

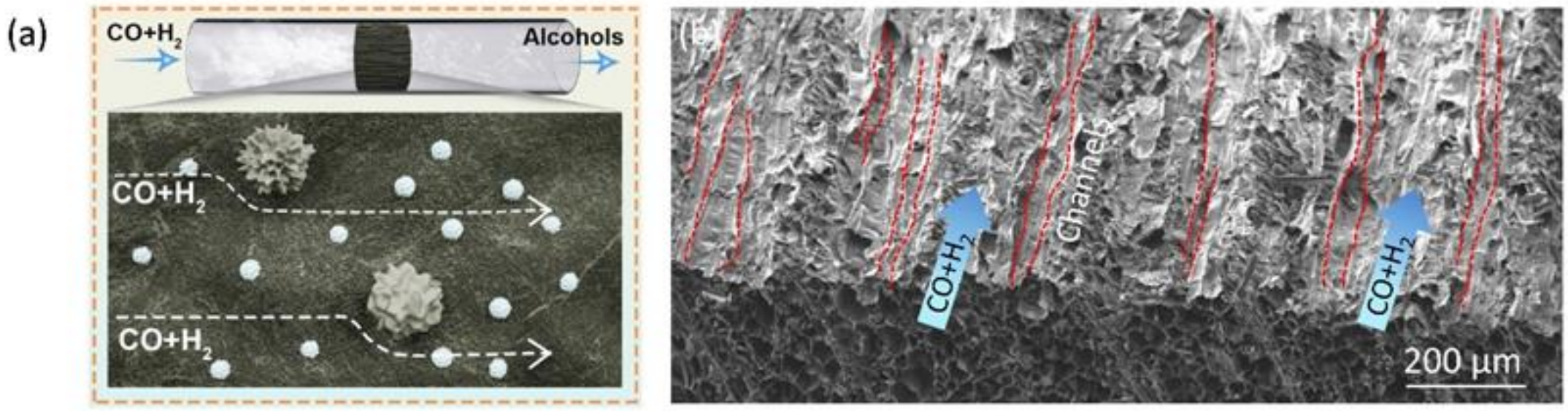

**Figure S21**. Schematic presentation of syngas thermal conversion into alcohols. (a) An illustrative image of syngas flow through carbon channels, where the catalytic reaction occurs. (b) A SEM image of the open 3D channels model, which ensures the smooth transport of CO and $H_2$ mixture.

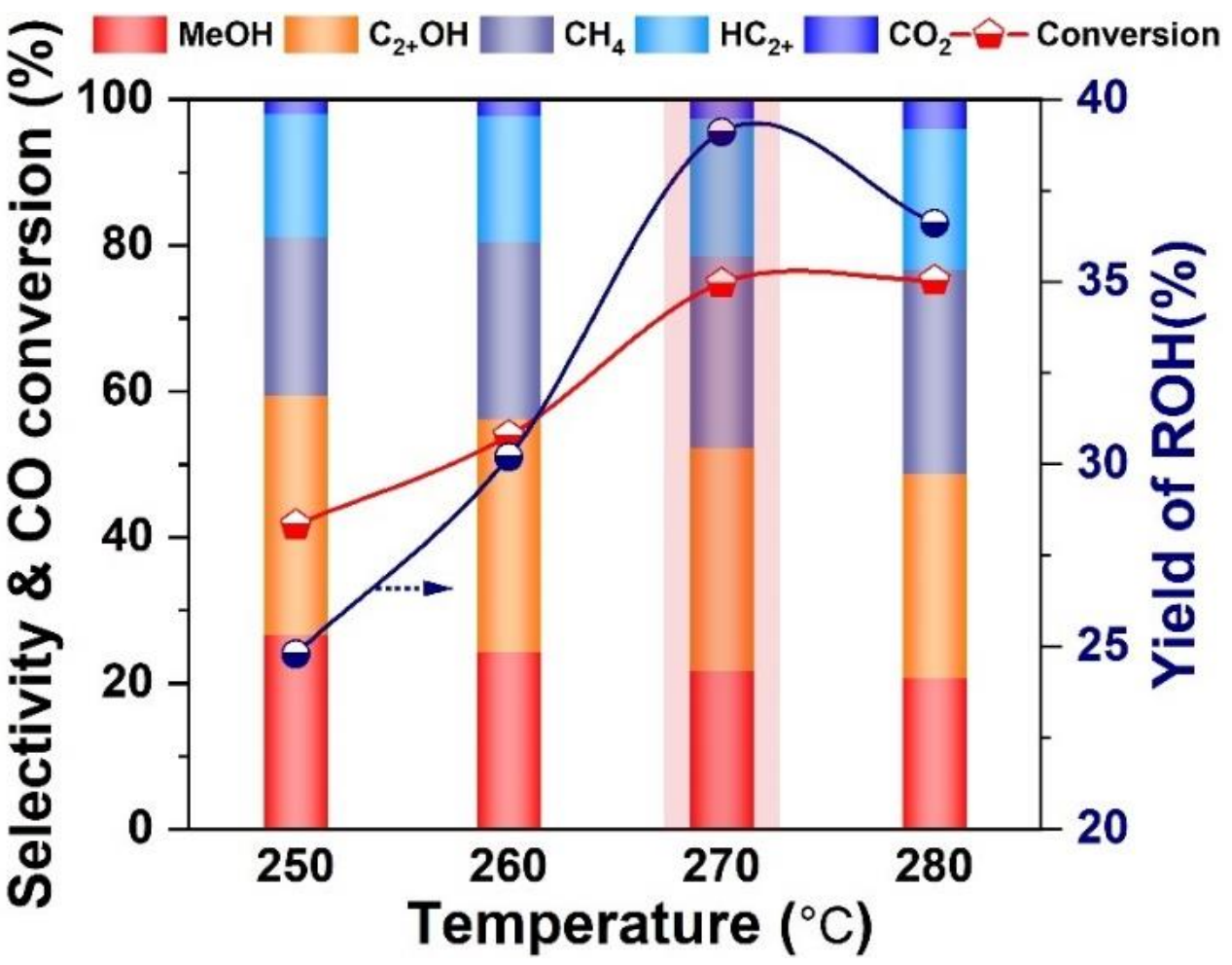

**Figure S22** The catalytic performance of Cu/Co-CW-$W_1P_1$ for HAS by changing the temperature.

The CO conversion is enhanced as increasing the reaction temperature, whilst decreasing the product selectivity of methanol and $C_{2+}OH$, accompanied by the increased selectivity of hydrocarbon and $CO_2$. The yield of alcohols obtained the highest value at 270 °C.

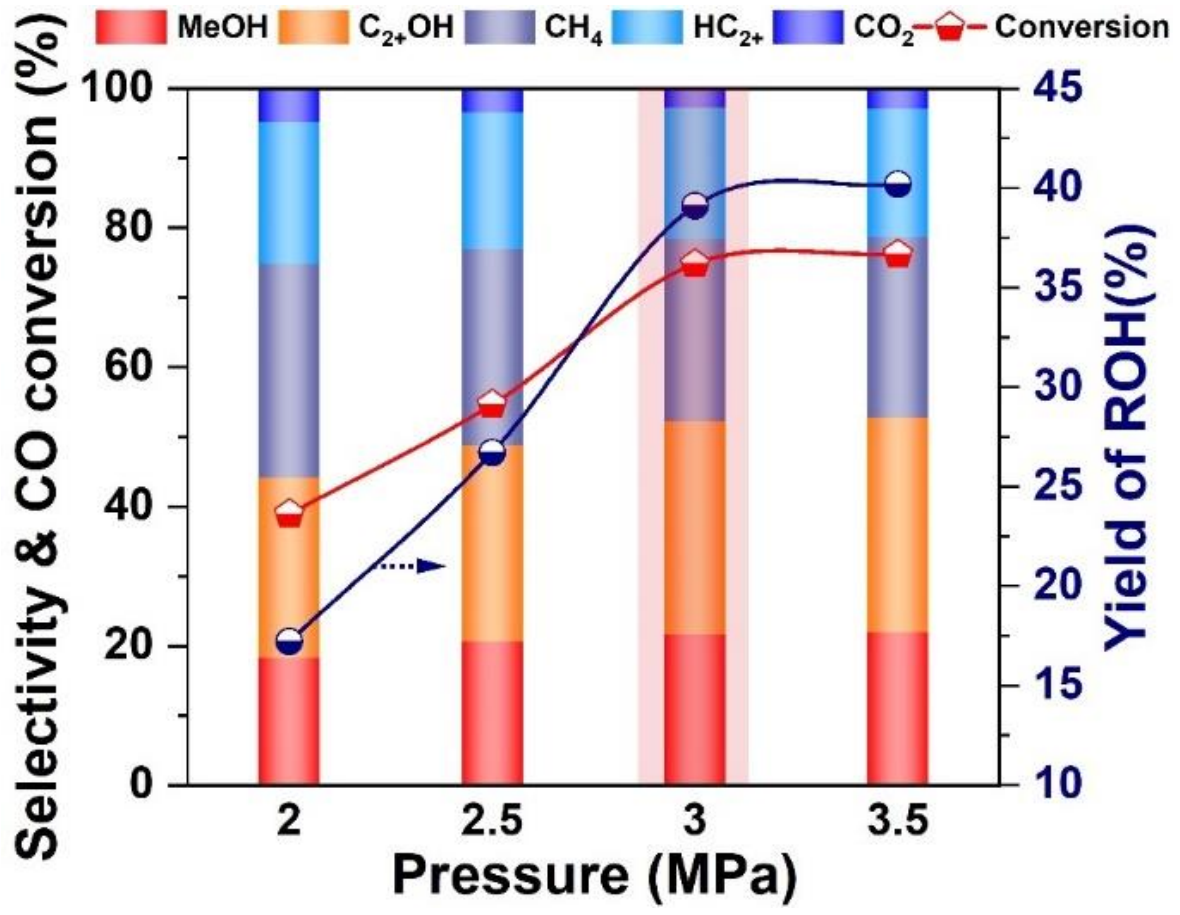

**Figure S23.** The catalytic performance of Cu/Co-CW-$W_1P_1$ for HAS over by changing the pressure.

The CO conversion increased as increasing reaction pressure, similar to the case for the methanol and $C_{2+}OH$. However, the selectivity to hydrocarbon and $CO_2$ decreased within the range of total reaction pressure from 2 MPa to 3.5 MPa. The optimal reaction pressure was 3MPa.

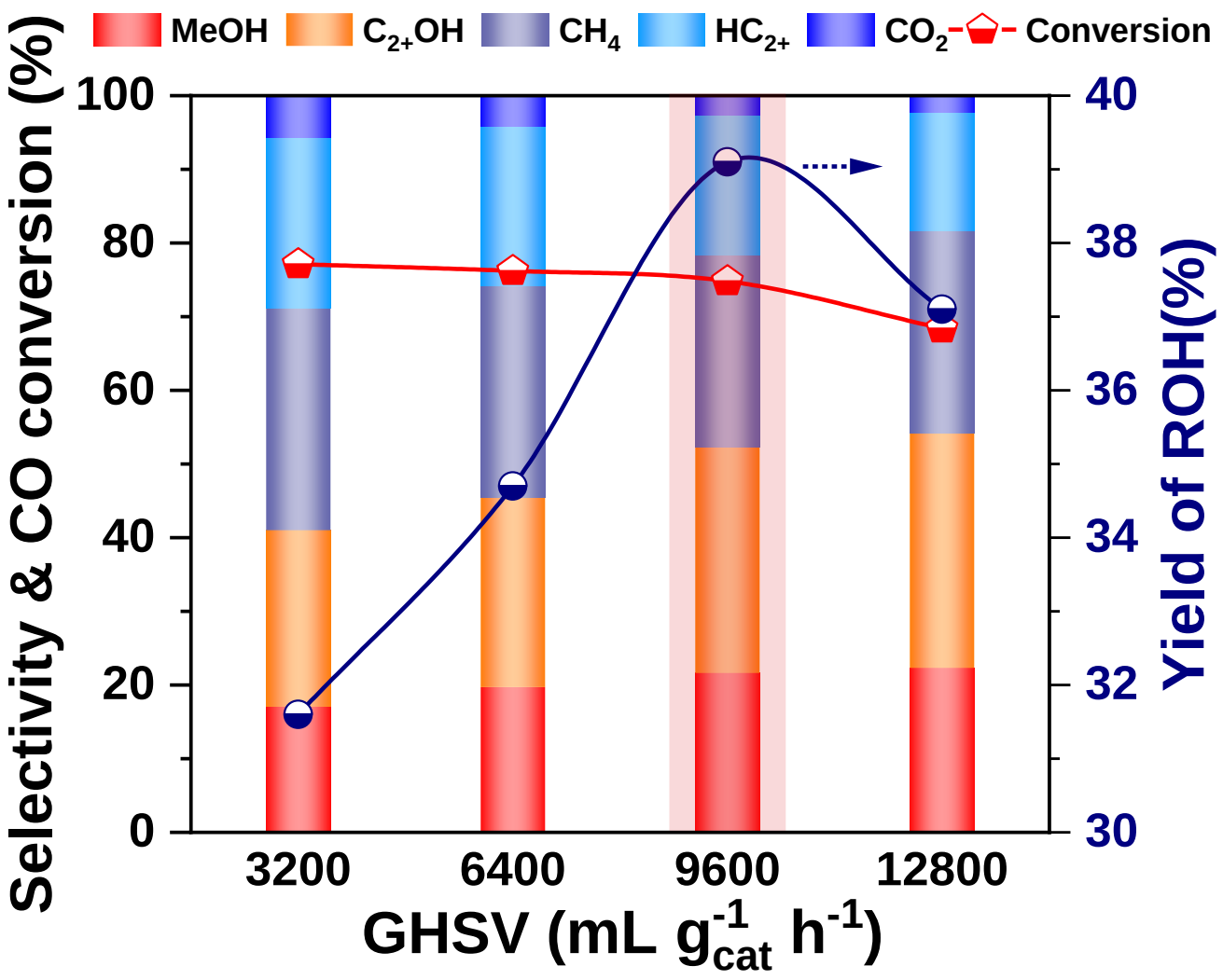

**Figure S24.** The catalytic performance Cu/Co-CW-$W_1P_1$ for HAS over by changing the gas hour space velocity (GHSV).

The CO conversion decreased as increasing the GHSV. As for the product distribution, an increase in the GHSV resulted in a decrease in $C_1$ by-product ($CH_4$ and $CO_2$), accompanied by a first increase and then decrease in methanol and $C_{2+}OH$. The optimal reaction GHSV was 9600 mL $g_{cat}^{-1}$ $h^{-1}$.

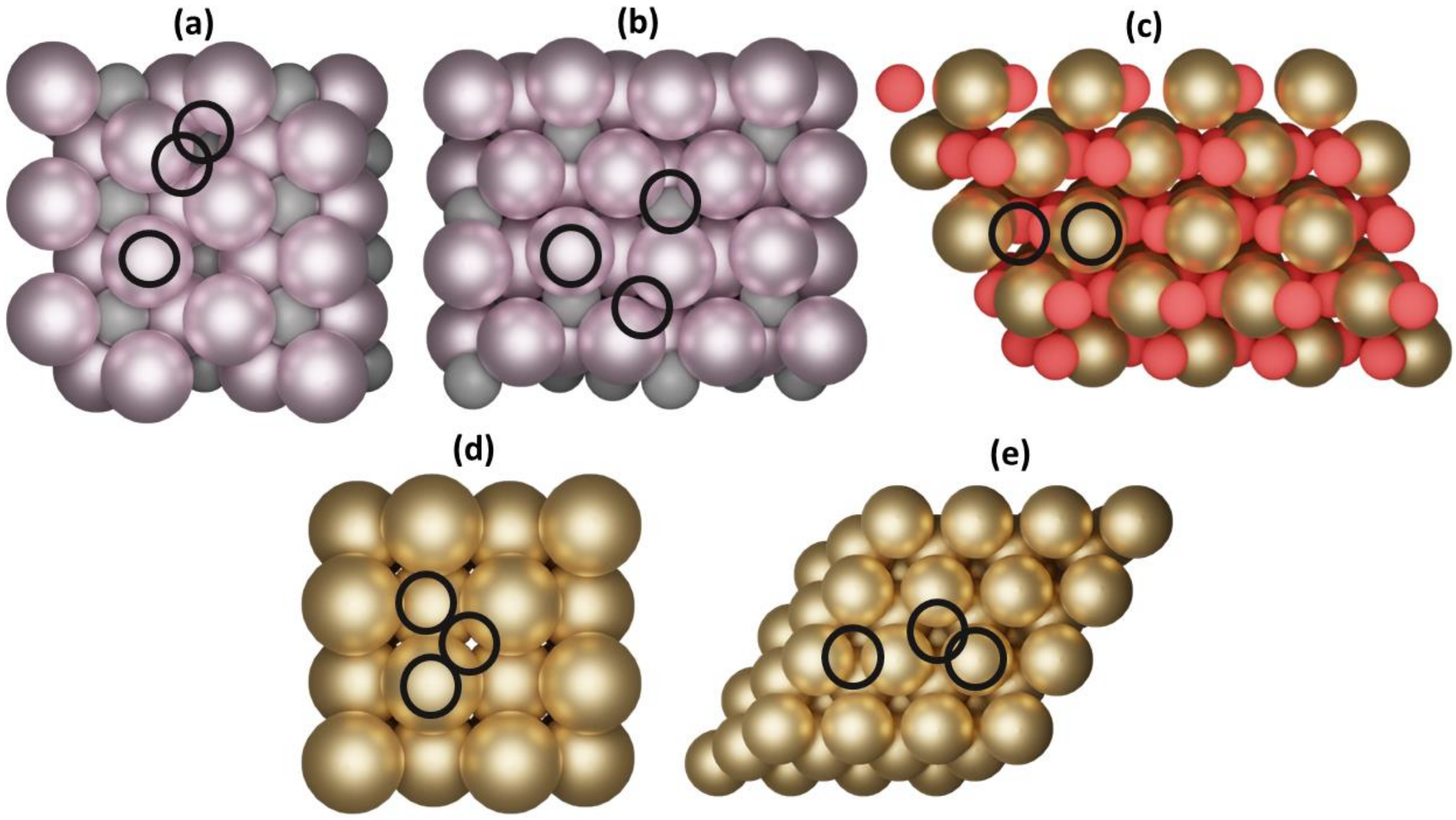

**Figure S25.** Studied adsorption sites on the (a) $Co_2C(020)$ and (b) $Co_2C(101)$, (c) CuO(111), (d) Cu(100), (e) Cu(111) surfaces from the top view.

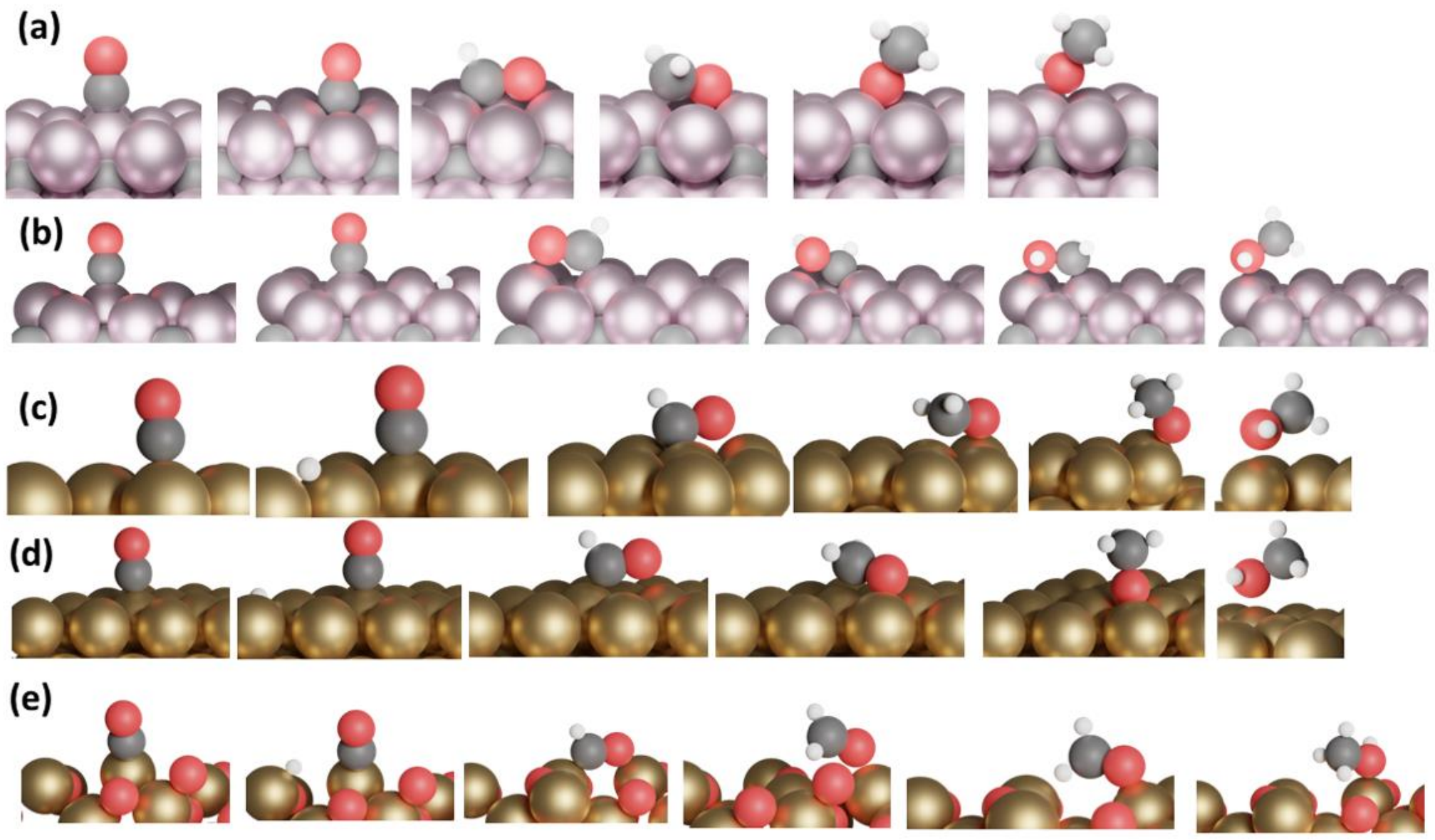

**Figure S26.** Optimal reaction mechanism toward MeOH over (a) $Co_2C$(020) and (b) $Co_2C$(101), (c) Cu(100), (d) Cu(111), and (e) CuO(111) surfaces from the side view.

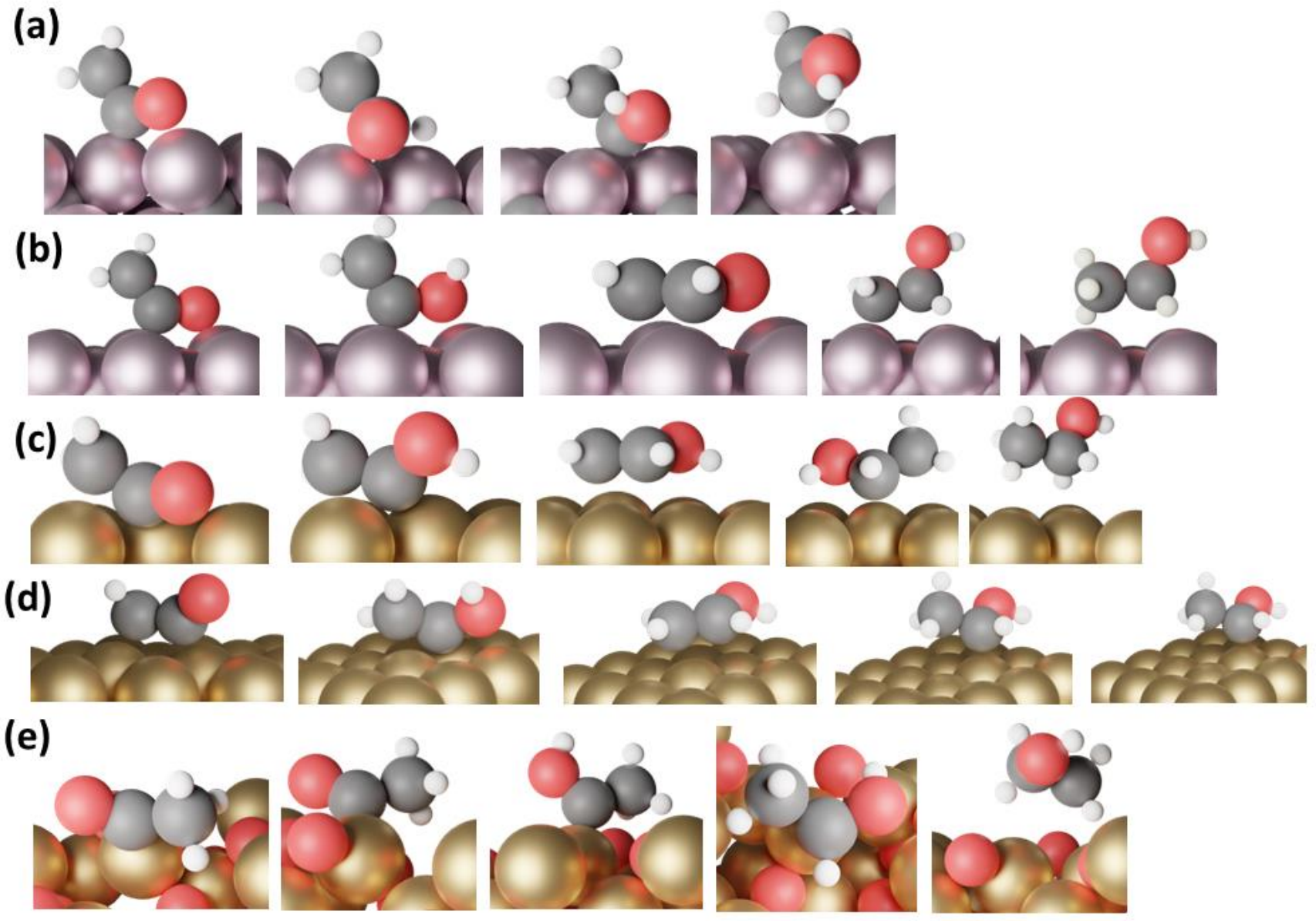

**Figure S27.** Optimal reaction mechanism toward EtOH over (a) $Co_2C$(020) and (b) $Co_2C$(101), (c) Cu(100), (d) Cu(111), and (e) CuO(111) surfaces from the side view.

**Table S1.** Preparation of the mixture solution for immersing natural wood (1,2-propanediol/deionized water mixture solution of 0.1 mol $L^{-1}$ $Co(NO_3)_2$•$6H_2O$ and 0.1 mol $L^{-1}$ $Cu(NO_3)_2$•$3H_2O$).

| Samples | Deionized water | 1,2-propanediol | Mixture (mol $L^{-1}$) | |
|---|---|---|---|---|
| | | | $Co(NO_3)_2$•$6H_2O$ | $Cu(NO_3)_2$•$3H_2O$ |
| Cu/Co-CW-W | 100 mL | 0 mL | 2.91 g | 2.42 g |
| Cu/Co-CW-$W_1P_1$ | 50 mL | 50 mL | 2.91 g | 2.42 g |
| Cu/Co-CW-$W_1P_2$ | 34 mL | 66 mL | 2.91 g | 2.42 g |
| Co-CW-$W_1P_1$ | 50 mL | 50 mL | 2.91 g | - |
| Cu-CW-$W_1P_1$ | 50 mL | 50 mL | - | 2.42 g |

**Table S2.** Catalyst performance of Cu/Co-CW catalysts for HAS.

| Catalyst | CO conv. (%) | Carbon selectivity (C mol%) | | | | |
|---|---|---|---|---|---|---|
| | | MeOH | $C_{2+}OH$ | $CH_4$ | $HC_{2+}$ | $CO_2$ |
| Cu/Co-CW-W | 42.9 | 21.4 | 13.7 | 31.7 | 25.3 | 7.9 |
| Cu/Co-CW-$W_1P_1$ | 74.8 | 21.6 | 30.7 | 26.1 | 18.9 | 2.7 |
| Cu/Co-CW-$W_1P_2$ | 61.2 | 19.3 | 23.6 | 29.3 | 23.2 | 4.6 |
| Co/CW-$W_1P_1$ | 23.7 | 12.6 | 13.0 | 26.9 | 16.4 | 31.1 |
| Cu/CW-$W_1P_1$ | 29.2 | 13.0 | 11.8 | 34.8 | 17.9 | 22.5 |

Reaction conditions: T=270 °C, GHSV=24000 mL $g_{cat}^{-1}$ $h^{-1}$, P=3.0 MPa, and $H_2$/CO=2/1, time on stream (TOS) = 48 h.

**Table S3.** Alcohol distribution of Cu/Co-CW catalysts in HAS.

| Catalyst | MeOH | EtOH | PrOH | BuOH | $C_{5+}OH$ |
|---|---|---|---|---|---|
| Cu/Co-CW-W | 61.6 | 26.2 | 8.4 | 3.1 | 1.2 |
| Cu/Co-CW-$W_1P_1$ | 41.3 | 40.7 | 9.9 | 5.8 | 2.3 |
| Cu/Co-CW-$W_1P_2$ | 44.9 | 38.5 | 9.4 | 5.2 | 2.0 |
| Co/CW-$W_1P_1$ | 49.2 | 36.6 | 9.5 | 3.2 | 1.5 |
| Cu/CW-$W_1P_1$ | 50.7 | 33.7 | 11.5 | 3.0 | 1.1 |

Reaction conditions: T=270 °C, GHSV=24000 mL $g_{cat}^{-1}$ $h^{-1}$, P=3.0 MPa, and $H_2/CO=2/1$, time on stream (TOS) = 48 h.

**Table S4**. The surface area, pore volume, and pore size of the CW, Cu/Co-CW-W, Cu/Co-CW-$W_1P_1$, and Cu/Co-CW- $W_1P_2$ correspond to Figure S4.

| Samples | Surface area ($m^2/g$) | Pore Volume ($cm^3/g$) | Pore Size (nm) |
|---|---|---|---|
| CW | 518 | 0.26 | 1.9 |
| Cu/Co-CW-W | 64 | 0.09 | 5.4 |
| Cu/Co-CW-$W_1P_1$ | 127 | 0.10 | 3.2 |
| Cu/Co-CW-$W_1P_2$ | 56 | 0.07 | 5.1 |

# 2. Supporting methodological details

## 2.1. Computational details

Gibbs free energy diagrams (**Tables S6-S9**) over $Co_2C(101)$, $Co_2C(020)$, Cu(100), Cu(111), and CuO(111) catalysts contain zero-point and entropic contributions (**Tables S10-S14**) calculated with PBE+D2 functional. Results were analyzed by thermochemistry formalism.[1] The intermediate's energies for Methanol (Eq. 1-11):

$$\Delta G^o_{DFT} = G(*CO) - G(*) - G(CO) \quad (1)$$
$$\Delta G^o_{DFT} = G(*CHO) - G(*CO*H) \quad (2)$$
$$\Delta G^o_{DFT} = G(*CHO) - G(*CO) + G(1/2H_2) \quad (3)$$
$$\Delta G^o_{DFT} = G(*COH) - G(*CO*H) \quad (4)$$
$$\Delta G^o_{DFT} = G(*COH) - G(*CO) + G(1/2H_2) \quad (5)$$
$$\Delta G^o_{DFT} = G(*OCH_2) - G(*CHO) + G(1/2H_2) \quad (6)$$
$$\Delta G^o_{DFT} = G(*CHOH) - G(*CHO) + G(1/2H_2) \quad (7)$$
$$\Delta G^o_{DFT} = G(*OCH_3) - G(*OCH_2) + G(1/2H_2) \quad (8)$$
$$\Delta G^o_{DFT} = G(*CH_2OH) - G(*CHOH) + G(1/2H_2) \quad (9)$$

[1]Ochterski, J. W.; Ph, D. Thermochemistry in Gaussian. Gaussian Inc Pittsburgh PA 2000, 264

$\Delta G^{o}_{DFT} = G(*CH_3OH) - G(*OCH_3) + G(1/2H_2)$ (10)

$\Delta G^{o}_{DFT} = G(*CH_3OH) - G(*CH_2OH) + G(1/2H_2)$ (11)

**Table S6**. Gibbs free energy diagrams of optimal reaction pathway for $CH_3OH$ production over $Co_2C(101)$. Gibss energies were calculated following the reaction equations above (**Eqs. 1-3, 7, 9,** and **11**).

| | $\Delta G_{Co_2C(101)}$ /eV |
|---|---|
| ***CO** | -1.57 |
| ***CO-H** | -0.60 |
| ***CHO** | -3.27 |
| ***CHOH** | 0.30 |
| ***$CH_2OH$** | -0.34 |
| ***$CH_3OH$** | -0.16 |
| **$CH_3OH_{(des)}$** | 4.38 |

**Table S7**. Gibbs free energy diagrams of optimal reaction pathway for $CH_3OH$ production over $Co_2C(020)$, Cu(100, 111), and CuO(111). The Gibbs energies were calculated following the reaction equations above (**Eqs. 1-3, 6, 8,** and **10**).

| | $\Delta G_{Co_2C(020)}$ /eV | $\Delta G_{Cu(100)}$ /eV | $\Delta G_{Cu(111)}$ /eV | $\Delta G_{CuO(111)}$ /eV |
|---|---|---|---|---|
| ***CO** | -1.69 | -0.42 | -0.33 | -0.81 |
| ***CO-H** | -0.85 | -0.01 | -0.10 | -0.33 |
| ***CHO** | 0.14 | 0.57 | 0.74 | -0.12 |
| ***$OCH_2$** | 0.10 | -0.14 | -0.26 | 0.42 |
| ***$OCH_3$** | -0.01 | -0.81 | -0.74 | 0.09 |
| ***$CH_3OH$** | 0.45 | -0.02 | -0.17 | -1.31 |
| **$CH_3OH_{(des)}$** | 0.15 | 0.48 | 0.50 | -0.02 |

The intermediate's energies for ethanol (**Eqs. 12-34**) were calculated as:

$\Delta G^{o}_{DFT} = G(*CH_3\ *O) - G(*CH_3O)$ (12)

$\Delta G^{o}_{DFT} = G(*CH_3*CO) - G(*CH_3) + G(CO)$ (13)

$\Delta G^{o}_{DFT} = G(*CH_3CO) - G(*CH_3*CO)$ (14)

$\Delta G^{o}_{DFT} = G(*CH_3CHO) - G(*CH_3CO) + G(1/2H_2)$ (15)

$\Delta G^{o}_{DFT} = G(*CH_3COH) - G(*CH_3CO) + G(1/2H_2)$ (16)

$\Delta G^{o}_{DFT} = G(*CH_3CHOH) - G(*CH_3CHO) + G(1/2H_2)$ (17)

$\Delta G^{o}_{DFT} = G(*CH_3CHOH) - G(*CH_3COH) + G(1/2H_2)$ (18)

$\Delta G^{o}_{DFT} = G(*CH_3CH_2OH) - G(*CH_3CHOH) + G(1/2H_2)$ (19)

$\Delta G^{o}_{DFT} = G(*CH_2*CO) - G(*CH_2) + G(CO)$ (20)

$\Delta G^{o}_{DFT} = G(*CH_2CO) - G(*CH_2*CO)$ (21)

$\Delta G^o_{DFT} = G(*CH_2*CO) - G(*CH_2) + G(CO)$ (22)

$\Delta G^o_{DFT} = G(*CH_3CO) - G(*CH_2CO) + G(1/2H_2)$ (23)

$\Delta G^o_{DFT} = G(*CH_2COH) - G(*CH_2CO) + G(1/2H_2)$ (24)

$\Delta G^o_{DFT} = G(*CH_3COH) - G(*CH_3CO) + G(1/2H_2)$ (25)

$\Delta G^o_{DFT} = G(*CH_3CHO) - G(*CH_3CO) + G(1/2H_2)$ (26)

$\Delta G^o_{DFT} = G(*CH_3COH) - G(*CH_2COH) + G(1/2H_2)$ (27)

$\Delta G^o_{DFT} = G(*CH_2CHOH) - G(*CH_2COH) + G(1/2H_2)$ (28)

$\Delta G^o_{DFT} = G(*CH_3CHOH) - G(*CH_3COH) + G(1/2H_2)$ (29)

$\Delta G^o_{DFT} = G(*CH_3CHOH) - G(*CH_3CHO) + G(1/2H_2)$ (30)

$\Delta G^o_{DFT} = G(*CH_3CHOH) - G(*CH_2CHOH) + G(1/2H_2)$ (31)

$\Delta G^o_{DFT} = G(*CH_2CH_2OH) - G(*CH_2CHOH) + G(1/2H_2)$ (32)

$\Delta G^o_{DFT} = G(*CH_3CH_2OH) - G(*CH_3CHOH) + G(1/2H_2)$ (33)

$\Delta G^o_{DFT} = G(*CH_3CH_2OH) - G(*CH_2CH_2OH) + G(1/2H_2)$ (34)

$\Delta G^o_{DFT} = G(CH_3CH_2OH_{(g)}) - G(*CH_3CH_2OH) + G(1/2H_2)$ (35)

**Table S8**. Gibbs free energy diagrams of optimal reaction pathway for $C_2H_5OH$ production over $Co_2C(101)$, Cu(100), and Cu(111). The Gibbs energies were calculated following the reaction equations above (**Eqs. 22, 24, 28, 31, 32, 33, 34, and 35**).

| | $\Delta G_{Co_2C(101)}$ /eV | $\Delta G_{Cu(100)}$ /eV | $\Delta G_{Cu(111)}$ /eV |
|---|---|---|---|
| **$*CH_2CO$** | -0.19 | -0.63 | -0.04 |
| **$*CH_2COH$** | 3.33 | 0.20 | -0.16 |
| **$*CH_2CHOH$** | -1.01 | -0.44 | -0.68 |
| **$*CH_2CH_2OH$** | -2.70 | - | - |
| **$*CH_3CHOH$** | - | 0.06 | 0.11 |
| **$*CH_3CH_2OH$** | -0.27 | -3.90 | -0.63 |
| **$CH_3CH_2OH_{(des)}$** | -1.10 | -1.29 | -1.30 |

**Table S9**. Gibbs free energy diagrams of optimal reaction pathway for $C_2H_5OH$ production over $Co_2C(020)$ and CuO(111). The Gibbs energies were calculated following the reaction equation above (**Eqs. 23, 25, 26, 29, 30, 33, and 35**).

| | $\Delta G_{Co_2C(020)}$ /eV | $\Delta G_{CuO(111)}$ /eV |
|---|---|---|
| **$*CH_2CO$** | - | -1.32 |
| **$*CH_3CO$** | -0.93 | -0.32 |
| **$*CH_3CHO$** | 0.44 | - |
| **$*CH_3COH$** | - | -3.66 |
| **$*CH_3CHOH$** | 3.39 | -0.18 |
| **$*CH_3CH_2OH$** | -3.38 | -3.17 |
| **$CH_3CH_2OH_{(des)}$** | -0.96 | -1.25 |

**Table S10**. Zero-point, and entropic (in eV) energies on the $Co_2C(020)$ surface at 543.15 K.

| | *ZPE* | $TS_{vib}$ |
|---|---|---|
| **$*CO$** | 0.37 | 0.30 |
| **$*CO\text{-}H$** | 0.55 | 0.41 |
| **$*CHO$** | 0.65 | 0.33 |
| **$*COH$** | 0.68 | 0.36 |
| **$*OCH_2$** | 0.98 | 0.37 |
| **$*OCH_3$** | 1.34 | 0.46 |
| **$*CH_2OH$** | 1.33 | 0.44 |
| **$*CH_3OH$** | 1.70 | 0.53 |
| **$*C$** | 0.16 | 0.11 |
| **$*CH$** | 0.46 | 0.18 |
| **$*CH_2$** | 0.74 | 0.25 |
| **$*CH_3$** | 1.06 | 0.33 |
| **$*CH_3\text{-}O$** | 1.22 | 0.48 |
| **$*CH_2\text{-}CO$** | 1.09 | 0.59 |
| **$*CH_3\text{-}CO$** | 1.42 | 0.68 |
| **$*CH_3CO$** | 1.53 | 0.57 |
| **$*CH_2CO$** | 1.12 | 0.52 |
| **$*CH_2COH$** | 1.51 | 0.54 |
| **$*CH_3COH$** | 1.85 | 0.61 |
| **$*CH_3CHO$** | 1.80 | 0.58 |
| **$*CH_2CHOH$** | 1.83 | 0.59 |
| **$*CH_3CHOH$** | 2.16 | 0.65 |
| **$*CH_2CH_2OH$** | 2.11 | 0.64 |
| **$*\ CH_3CH_2OH$** | 2.48 | 0.72 |

**Table S11**. Zero-point, and entropic (in eV) energies on the $Co_2C(101)$ surface at 543.15 K.

| | *ZPE* | $TS_{vib}$ |
|---|---|---|
| ***CO** | 0.37 | 0.29 |
| ***CO-H** | 0.58 | 0.35 |
| ***CHO** | 0.66 | 0.35 |
| ***COH** | 0.68 | 0.35 |
| ***$OCH_2$** | 1.00 | 0.47 |
| ***CHOH** | 1.00 | 0.39 |
| ***$OCH_3$** | 1.28 | 0.45 |
| ***$CH_2OH$** | 1.35 | 0.43 |
| ***$CH_3OH$** | 1.71 | 0.52 |
| ***C** | 0.16 | 0.11 |
| ***CH** | 0.46 | 0.17 |
| ***$CH_2$** | 0.77 | 0.27 |
| ***$CH_3$** | 1.10 | 0.34 |
| ***$CH_3$-O** | 1.25 | 0.47 |
| ***$CH_2$-CO** | 1.12 | 0.52 |
| ***$CH_3$-CO** | 1.47 | 0.62 |
| ***$CH_2CO$** | 1.18 | 0.53 |
| ***$CH_3CO$** | 1.53 | 0.57 |
| ***$CH_2COH$** | 1.53 | 0.59 |
| ***$CH_3COH$** | 1.24 | 0.63 |
| ***$CH_3CHO$** | 2.48 | 0.60 |
| ***$CH_2CHOH$** | 1.86 | 0.55 |
| ***$CH_2CH_2OH$** | 2.16 | 0.66 |
| ***$CH_3CH_2OH$** | 2.48 | 0.69 |

**Table S12**. Zero-point, and entropic (in eV) energies on the CuO(111) surface at 543.15 K.

| | *ZPE* | $TS_{vib}$ |
|---|---|---|
| **$*CO$** | 0.38 | 0.32 |
| **$*CO-H$** | 0.61 | 0.38 |
| **$*CHO$** | 0.67 | 0.35 |
| **$*OCH_2$** | 1.02 | 0.44 |
| **$*OCH_3$** | 1.29 | 0.42 |
| **$*CH_3OH$** | 1.72 | 0.55 |
| **$*CH$** | 0.46 | 0.20 |
| **$*CH_2$** | 0.79 | 0.24 |
| **$*CH_3$** | 1.15 | 0.33 |
| **$*CH_3-O$** | 1.30 | 0.49 |
| **$*CH_3-CO$** | 1.52 | 0.64 |
| **$*CH_2-CO$** | 1.18 | 0.54 |
| **$*CH_3CO$** | 1.55 | 0.58 |
| **$*CH_2CO$** | 1.20 | 0.51 |
| **$*CH_3COH$** | 1.84 | 0.60 |
| **$*CH_2COH$** | 1.54 | 0.62 |
| **$*CH_3CHO$** | 1.88 | 0.65 |
| **$*CH_3CHOH$** | 2.22 | 0.65 |
| **$*CH_2CHOH$** | 1.89 | 0.58 |
| **$*CH_2CH_2OH$** | 2.19 | 0.62 |
| **$*CH_3CH_2OH$** | 2.57 | 0.74 |

**Table S13**. Zero-point, and entropic (in eV) energies on the Cu(100) surface at 543.15 K.

| | *ZPE* | $TS_{vib}$ |
|---|---|---|
| **$*CO$** | 0.37 | 0.33 |
| **$*CO$-H** | 0.58 | 0.43 |
| **$*CHO$** | 0.65 | 0.38 |
| **$*COH$** | 0.67 | 0.39 |
| **$*OCH_2$** | 0.98 | 0.44 |
| **$*CHOH$** | 1.01 | 0.40 |
| **$*OCH_3$** | 1.34 | 0.46 |
| **$*CH_2OH$** | 1.35 | 0.46 |
| **$*CH_3OH$** | 1.71 | 0.55 |
| **$*C$** | 0.16 | 0.12 |
| **$*CH$** | 0.46 | 0.18 |
| **$*CH_2$** | 0.77 | 0.27 |
| **$*CH_3$** | 1.11 | 0.36 |
| **$*CH_3$-O** | 1.26 | 0.53 |
| **$*CH_3$-CO** | 1.48 | 0.69 |
| **$*CH_2$-CO** | 1.15 | 0.61 |
| **$*CH_3CO$** | 1.53 | 0.59 |
| **$*CH_2CO$** | 1.18 | 0.50 |
| **$*CH_3COH$** | 1.86 | 0.64 |
| **$*CH_2COH$** | 1.52 | 0.57 |
| **$*CH_3CHO$** | 1.86 | 0.66 |
| **$*CH_3CHOH$** | 2.20 | 0.67 |
| **$*CH_2CHOH$** | 1.88 | 0.61 |
| **$*CH_2CH_2OH$** | 2.18 | 0.68 |
| **$*CH_3CH_2OH$** | 2.55 | 0.75 |

**Table S14**. Zero-point, and entropic (in eV) energies on the Cu(111) surface at 543.15 K.

| | *ZPE* | $TS_{vib}$ |
|---|---|---|
| **$*CO$** | 0.37 | 0.33 |
| **$*CO\text{-}H$** | 0.57 | 0.38 |
| **$*CHO$** | 0.66 | 0.38 |
| **$*COH$** | 0.68 | 0.39 |
| **$*OCH_2$** | 0.99 | 0.42 |
| **$*CHOH$** | 1.02 | 0.42 |
| **$*OCH_3$** | 1.36 | 0.47 |
| **$*CH_2OH$** | 1.33 | 0.47 |
| **$*CH_3OH$** | 1.42 | 0.56 |
| **$*C$** | 0.16 | 0.11 |
| **$*CH$** | 0.46 | 0.18 |
| **$*CH_2$** | 0.76 | 0.31 |
| **$*CH_3$** | 1.11 | 0.36 |
| **$*CH_3\text{-}O$** | 1.26 | 0.50 |
| **$*CH_3\text{-}CO$** | 1.49 | 0.70 |
| **$*CH_2\text{-}CO$** | 1.13 | 0.65 |
| **$*CH_3CO$** | 1.54 | 0.60 |
| **$*CH_2CO$** | 1.18 | 0.54 |
| **$*CH_3COH$** | 1.87 | 0.63 |
| **$*CH_2COH$** | 1.53 | 0.58 |
| **$*CH_3CHO$** | 1.87 | 0.66 |
| **$*CH_3CHOH$** | 2.21 | 0.68 |
| **$*CH_2CHOH$** | 1.89 | 0.61 |
| **$*CH_2CH_2OH$** | 2.20 | 0.66 |
| **$*CH_3CH_2OH$** | 2.55 | 0.75 |